\documentclass[9pt,twocolumn]{extarticle}

\usepackage{amsfonts}
\usepackage[dvips]{graphicx}
\usepackage{color}
\usepackage{amsthm,amsmath,amssymb}
\usepackage{mathrsfs}
\usepackage{epstopdf}
\usepackage{dsfont}
\usepackage{setspace}
\usepackage{multirow}
\usepackage{booktabs}
\usepackage{tabularx}
\usepackage{adjustbox}
\usepackage{colortbl}
\usepackage{hhline}
\usepackage{xcolor}
\usepackage{babel}
\usepackage{float} 
\usepackage{dblfloatfix}
\usepackage{xurl} 
\usepackage{hyperref}
\usepackage[numbers,sort&compress]{natbib}
\usepackage[normalem]{ulem}

\usepackage{braket}
\usepackage{mathtools}
\usepackage{hhline}
\definecolor{nred}{rgb}{0,0,0}
\newcommand{\red}{\textcolor{nred}}
\allowdisplaybreaks[4]
\providecommand{\bibcommenthead}{}
\providecommand{\bibinfo}[2]{#2}

\def\mytitle{
Classifying Multipartite Continuous Variable Entanglement Structures through Data-augmented Neural Networks
\vspace{-2mm}}      
 
\title{\vspace{-1.cm}\huge\textbf{\textrm{\mytitle}}}  

\author{Xiaoting~Gao$^{1,9}$, Mingsheng~Tian$^{1,8,9}$, Feng-Xiao~Sun$^{1,2}$, Ya-Dong~Wu$^{3}$, Yu~Xiang$^{4,*}$, Qiongyi~He$^{1,5,6,7,*}$}
\date{}

\begin{document}
\addtocontents{toc}{\protect\setcounter{tocdepth}{-1}}
\twocolumn[{
\maketitle 
\vspace{-9mm}
\begin{center}
\begin{minipage}{1\textwidth}
\begin{center}
\textit{\textrm{
\textsuperscript{1} State Key Laboratory of Artificial Microstructure and Mesoscopic Physics, School of Physics, Frontiers Science Center for Nano-optoelectronics, \& Collaborative Innovation Center of Quantum Matter, Peking University, Beijing 100871, China
\\\textsuperscript{2} State Key Laboratory of Information Photonics and Optical Communications \& School of Physical Science and Technology, Beijing University of Posts and Telecommunications, Beijing 100876, China
\\\textsuperscript{3} John Hopcroft Center for Computer Science, Shanghai Jiao Tong University, Shanghai 200240, China
\\\textsuperscript{4} Ministry of Education Key Laboratory for Nonequilibrium Synthesis and Modulation of Condensed Matter, Shaanxi Province Key Laboratory of Quantum Information and Quantum Optoelectronic Devices, School of Physics, Xi'an Jiaotong University, Xi'an 710049, China
\\\textsuperscript{5} Collaborative Innovation Center of Extreme Optics, Shanxi University, Taiyuan, Shanxi 030006, China
\\\textsuperscript{6} Peking University Yangtze Delta Institute of Optoelectronics, Nantong, Jiangsu 226010, China
\\\textsuperscript{7} Hefei National Laboratory, Hefei 230088, China
\\\textsuperscript{8} Present address: Department of Physics, The Pennsylvania State University, Pennsylvania, USA
\\\textsuperscript{9} These authors contributed equally: Xiaoting~Gao and Mingsheng~Tian
\\\textsuperscript{*}e-mail: xiangy@xjtu.edu.cn; qiongyihe@pku.edu.cn
}}
\end{center}
\end{minipage}
\end{center}

\setlength\parindent{12pt}
\begin{quotation}
\noindent 
{\bf{Neural networks have emerged as a promising paradigm for quantum information processing, yet they confront the challenge of generating training datasets with sufficient size and rich diversity, which is particularly acute when dealing with multipartite quantum systems.
For instance, in the task of classifying different structures of multipartite entanglement in continuous variable systems, it is necessary to simulate a large number of infinite-dimensional state data that can cover as many types of non-Gaussian states as possible.
Here, we develop a data-augmented neural network to address this task with homodyne measurement data. 
A quantum data augmentation method based on classical data processing techniques and quantum physical principles is proposed to efficiently enhance network performance. 
By testing on randomly generated tripartite and quadripartite states, we demonstrate that the network can infer the entanglement structure among the various partitions, and the accuracies are significantly improved with data augmentation. 
Our approach allows us to further extend the use of data-driven machine learning techniques to more complex tasks of learning quantum systems encoded in a large Hilbert space.
}} 
\end{quotation}}] 

\section{Introduction}
Inspired by biological systems, neural networks have demonstrated remarkable success across diverse domains including quantum physics. 
These models offer innovative solutions to challenges such as quantum state reconstruction~\cite{Lohani_2020,PhysRevLett.127.140502,ahmed2021classification}, feature learning~\cite{cimini2020neural,zhu2022flexible,gebhart2023learning,PhysRevX.12.011059,PhysRevLett.130.210601,gao2023correlation}, and even fundamental physics discovery~\cite{PhysRevLett.121.111801,PhysRevLett.121.241803,PhysRevLett.124.010508,karagiorgiMachine2022}.
However, the efficacy of these data-driven techniques—for both classical and quantum problems—depends critically on the quality of the underlying datasets~\cite{Jain2020,Gupta2021,GONG2023107268}.
While increasing the size and diversity of training data enhances a network's ability to generalize to unseen cases, assembling such comprehensive datasets demands substantial time and resources for meticulous collection and annotation.
\par
\begin{figure*}[h]
    \centering
    \includegraphics[width=1\linewidth]{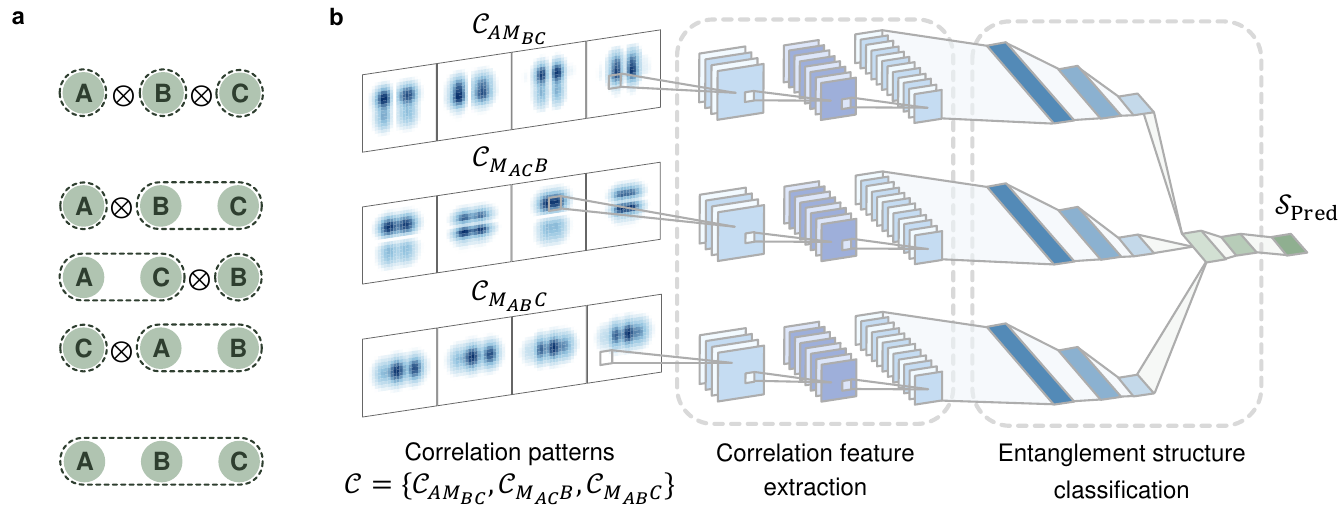}
    \caption{\textbf{The multi-class classification task.} \textbf{a,} The tripartite entanglement structures for the partitions of fully separable partition $1\otimes1\otimes1$, three biseparable partitions $1\otimes2$, and fully inseparable states.
    \textbf{b,} The neural network architecture for this task. Three groups of correlation patterns $\mathcal{C}_{AM_{BC}}$, $\mathcal{C}_{M_{AC}B}$, and $\mathcal{C}_{M_{AB}C}$ provide correlation features of $A-BC$, $B-AC$, and $C-AB$ splittings, respectively. $M_{ij}$ $(i,j \in \{A, B, C\})$ denotes one of the output modes by mixing modes $i$ and $j$ on a balanced beam splitter. This mixing strategy enables more effective extraction of information across different partitions while reducing the number of required measurements. The features are extracted by convolutional layers and subsequently classified by fully connected layers into three classes of tripartite entanglement structures as shown in \textbf{a}, resulting in a predicted entanglement structure label $\mathcal{S}_{\text{Pred}}$.}
    \label{fig:network}
\end{figure*}

This becomes more challenging when dealing with multipartite quantum systems, where complexity escalates with the number of subsystems. Detecting multipartite entanglement exemplifies this difficulty. Multipartite entanglement, essential in quantum physics, is classified by its separability across all possible combinations of subsystems, i.e., multi-partitions~\cite{RevModPhys.80.517,RevModPhys.81.865,GUHNE20091,PhysRevLett.111.110503}. 
For some typical states, such as multipartite Gaussian states, entanglement conditions in terms of different multi-partitions have been demonstrated~\cite{comb}. 
However, classifying the entanglement structures of arbitrary multipartite non-Gaussian states, whose nontrivial correlations exist in higher-order moments, remains experimentally infeasible. 
\par

Efforts to harness neural networks for classifying entanglement structures in discrete variable systems have been made~\cite{grayMachine2018,chenEntanglement2021,chenDetecting2022,chenCertify2023}, aiming to circumvent the high cost of full quantum state tomography~\cite{PhysRevLett.107.020404}. Conceptually, this approach aligns with broader machine-learning studies that extract global quantum properties from high-dimensional many-body data. For example, classical neural networks have been used to identify phases of matter from spin configurations and entanglement spectra~\cite{nphys4035,nphys4037}, and quantum convolutional neural networks process quantum states directly through structured quantum circuits~\cite{s41567-019-0648-8}. In contrast to these finite-dimensional many-body settings, the present work addresses multipartite entanglement structure classification in continuous variable (CV) systems. In such systems, characterized by infinite-dimensional Hilbert spaces~\cite{RevModPhys.77.513}, conventional tomography becomes even less practical~\cite{lvovsky2009continuous}.
Neural networks have offered an experimentally viable alternative by analyzing statistical features extracted from quadrature components of light through homodyne measurements, demonstrating success in bipartite settings~\cite{gao2023correlation}. 
For multipartite non-Gaussian states, however, the key difficulty lies not only in designing an effective classifier, but also in overcoming the major obstacle of generating sufficiently diverse and reliably labelled training samples--a challenge that hinders the direct generalization of existing techniques to this setting.
\par

The contributions of this work are twofold. First, we develop a neural network model capable of accurately classifying entanglement structures of arbitrary multipartite CV states using experimentally accessible homodyne measurement data as input. 
Second, to address the challenge in simulating multipartite CV systems, we propose a quantum data augmentation (QDA) method that significantly enhances the accuracy of predicting entanglement structures. 
Data augmentation refers to expanding datasets by transforming given data into new, label-preserving samples. In the quantum context, these operations can be derived from quantum physical principles such as entanglement invariance under mode permutation and the convexity of separable states. 
\par

We apply the data-augmented neural network for tripartite and quadripartite CV systems in all possible multi-partitions. The dataset covers states with varying degrees of non-Gaussian complexity, including highly non-Gaussian cases such as cat states. Compared with accuracies of $0.961$ and $0.796$ obtained using only original datasets for the two cases, the network with augmented data can remarkably improve them to $0.986$ and $0.928$, respectively.
Therefore, our results provide an efficient approach for classifying multipartite non-Gaussian entanglement structures in an experimentally accessible way, which could exhibit metrological power in quantum phase estimation~\cite{metro}.
Moreover, inspired by the resource-preserving free operations in quantum resource theory~\cite{RevModPhys.91.025001}, a wider range of label-preserving transformations can be designed as QDA operations. 
Hence, neural networks can be extended for broader tasks in learning multipartite quantum systems, even when data acquisition cost is exorbitant.
\par

\section{Results}
Multipartite entanglement detection goes beyond simply identifying the presence of entanglement; it also involves determining its detailed structure. 
Specifically, any $\mathcal{R}_1|\cdots|\mathcal{R}_j$-separable state can be decomposed in the form $\hat\rho=\sum_\gamma p_\gamma\hat\rho_{\mathcal{R}_1}^{(\gamma)}\otimes\cdots\otimes\hat\rho_{\mathcal{R}_j}^{(\gamma)}$, where 
$\mathcal{R}_i$ describes an ensemble of modes and $\hat\rho_{\mathcal{R}_i}^{(\gamma)}$ denotes the reduced density matrix of $\hat\rho^{(\gamma)}$ on the subsystem $\mathcal{R}_i$~\cite{Gessner2017entanglement}. In the following task, we classify cases with the same number of modes in each subsystem $\mathcal{R}_i$ as the same multi-partition class and label them by $\mathcal{S}$.
For example, states with separability $A|BC$, $B|AC$, and $C|AB$ all belong to the multi-partition class $\mathcal{S}=1\otimes2$.

\subsection{Quantum state set construction}
We train the neural network on a broad set of random mixed CV states, including all Gaussian states and most experimentally achievable non-Gaussian states. 
As illustrated in Fig.~\ref{fig:network}(a), we start with the simplest multipartite case: tripartite states.
To collect balanced samples for all possible classes of states with different multi-partitions $\mathcal{S}$ in a random way, we first generate a large number of $m$-mode ($m\leq 3$) seed states $\hat\sigma_m$. Then we use these seed states to separately construct each class.
For example, states with $\mathcal{S}=1\otimes1\otimes1$ can be decomposed into a tensor product of three single-mode seed states $\otimes_{i=1}^3\hat\sigma_1^{(i)}$, while states with $\mathcal{S}=1\otimes2$ can be decomposed into a tensor product $\hat\sigma_1\otimes\hat\sigma_2$ of a single-mode seed state $\hat\sigma_1$ and a two-mode entangled seed state $\hat\sigma_2$.\par

The seed states $\hat\sigma_m$ are generated based on the stellar formalism, which gives an operational characterization of non-Gaussian states through their stellar functions~\cite{Chabaud2020stellar,Chabaud2021classical,Chabaud2022holomorphic}. 
For a normalized pure state $|\psi\rangle$, the stellar function is defined as $F_\psi^{\star}(\mathbf{z}) \equiv \text{e}^{\frac{1}{2}\|\mathbf{z}\|^2}\left\langle \mathbf{z}^*|\psi\right\rangle$, where $|\mathbf{z}\rangle$ denotes a coherent state with complex amplitude $\mathbf{z}=(z_1,\ldots,z_m)\in \mathbb{C}^m$. Coherent states are the eigenstates of the annihilation operator and serve as the building blocks of CV quantum optics; their complex amplitudes parameterize displacements in phase space. 
The number of zeros of $F_\psi^{\star}(\mathbf{z})$ defines the stellar rank $r$, which quantifies the non-Gaussianity of state $|\psi\rangle$ and represents the minimal number of non-Gaussian operations required to engineer the state from the vacuum. Gaussian states, including coherent, squeezed, and thermal states, are formally characterized by Gaussian characteristic functions and quasi-probability distributions on the multimode phase space; consequently, all Gaussian states have $r=0$, and the stellar rank is preserved under all Gaussian unitaries. This formalism thus provides a rigorous yet accessible way to classify non-Gaussian resources in CV quantum systems. 

We focus on states with $r\in\{0,1,2,3,+\infty\}$, which include both finite- and infinite-stellar-rank states. For the finite case, any $m$-mode pure state can be decomposed into $\hat G|C(r)\rangle$, where $\hat G$ is a general $m$-mode Gaussian unitary consisting of squeezing, displacement, and beam splitters, and $|C(r)\rangle$, called a core state~\cite{Chabaud2021classical}, is an $m$-mode state with finite support over the Fock basis and stellar rank $r$. 
To summarize, each $m$-mode seed state $\hat\sigma_m$ is obtained from a random core state $|C(r)\rangle$ with a given finite stellar rank $r$, followed by an $m$-mode Gaussian unitary with random parameters and $m$ loss channels~\cite{Eaton2022measurementbased} with random efficiencies $\eta_l$ ($l=1, \dots, m$), given by 
\begin{align}
\hat\sigma_m=\left(\prod_{l=1}^m\hat L_l(\eta_l)\right)\hat G |C(r)\rangle\langle C(r)|\hat G^\dagger \left(\prod_{l=1}^m\hat L_l^\dagger(\eta_l)\right),
\end{align}
where each $\hat{L}_l$ is a Kraus operator of the loss channel~\cite{ou1992realization,LossChannelLect}. For the infinite case, seed states consist of multimode entangled cat states and single-mode coherent states, subjected to the same loss channels (Supplementary I.A and B).
\par
\clearpage
\begin{figure*}[h]
    \centering
    \includegraphics[width=0.95\linewidth]{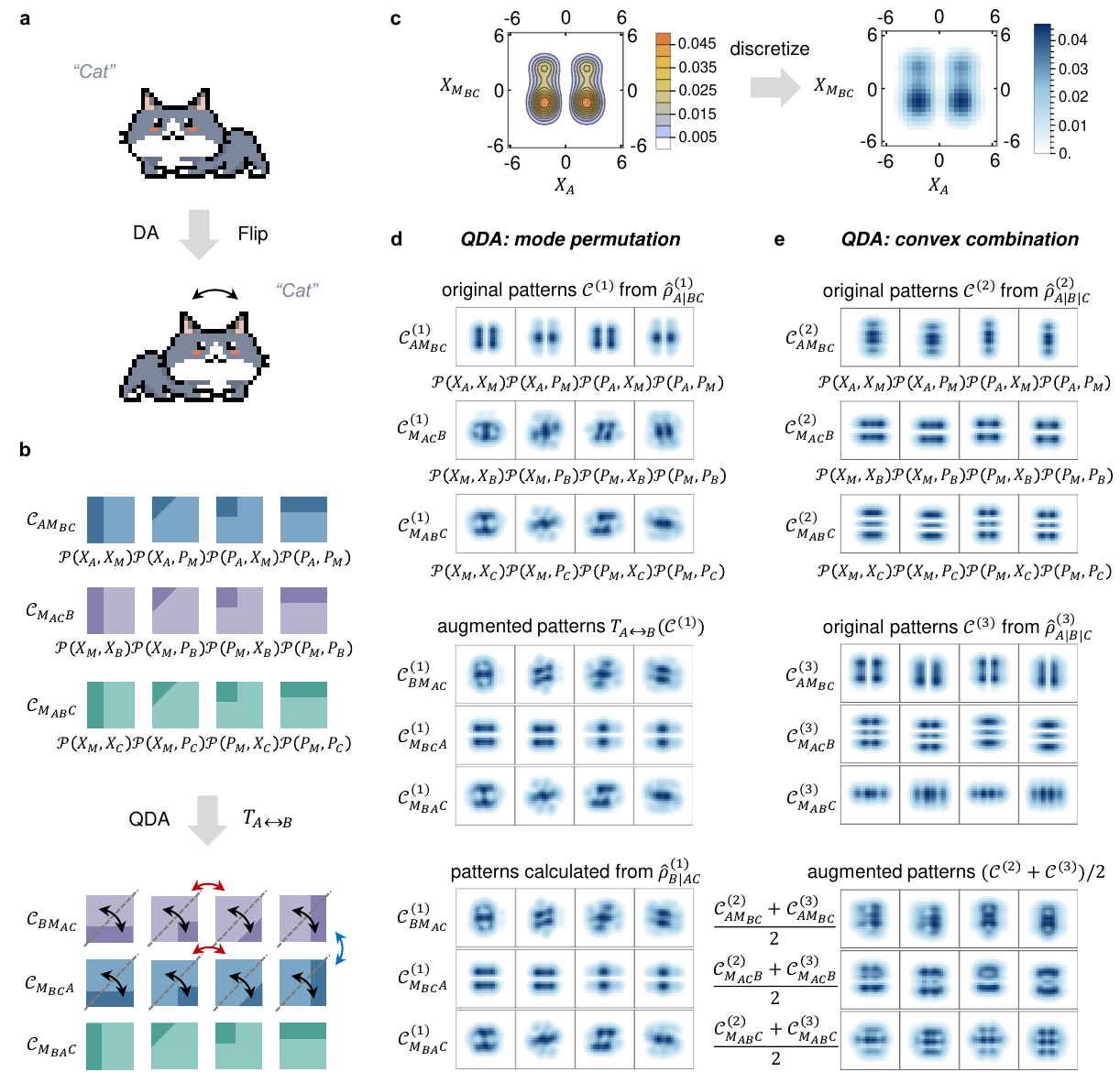}
    \caption{\textbf{Possible strategies of QDA on the input correlation patterns.}
    \textbf{a,} In the classical image augmentation process, flipping creates new samples without altering the semantic label ``cat''.
    \textbf{b,} Schematic of how correlation patterns $\mathcal{C}$ are transformed when modes $A$ and $B$ are swapped, while maintaining both the validity of transformed quantum measurement data and the invariance of the multi-partitions label $\mathcal{S}$. Black arrows represent flipping along the dashed lines, red arrows represent exchanging the patterns, and the blue arrow represents swapping two rows.
    \textbf{c,} Left: Theoretical joint probability distribution $\mathcal{P}(X_A,X_{M_{BC}})$ of a tripartite non-Gaussian state (Eq.~\ref{eq:distribution}). Right: The corresponding correlation pattern obtained by binning the distribution shown on the left.
    \textbf{d,} Correlation patterns $\mathcal{C}^{(1)}$ of an original $A|BC$ biseparable state $\hat\rho_{A|BC}^{(1)}$ (top).  By rearranging $\mathcal{C}^{(1)}$ according to the mode-permutation transformation $T_{A\leftrightarrow B}$ illustrated in \textbf{b}, QDA efficiently produces the corresponding patterns for the $B|AC$ partition at negligible time cost (middle). The same patterns can be obtained by explicitly constructing the permuted state $\hat\rho_{B|AC}^{(1)}$ and then recalculating the correlation patterns from its density matrix (bottom), but this direct simulation route is substantially more time-consuming.
    \textbf{e,} Correlation patterns $\mathcal{C}^{(2)}$ and $\mathcal{C}^{(3)}$ of two distinct fully separable states $\hat\rho_{A|B|C}^{(2)}$ and $\hat\rho_{A|B|C}^{(3)}$ (top and middle). A new sample of patterns  can be generated via QDA through the convex combination $(\mathcal{C}^{(2)}+\mathcal{C}^{(3)})/2$ (bottom).}
    \label{fig:aug}
\end{figure*}
\clearpage
Following the definition of entanglement structure in terms of multi-partitions, 
entangled seed states are required as necessary components in the cases except for fully separable partitions. 
Furthermore, these entangled seed states must be fully inseparable to rule out any possible separability from seed states and avoid overlap between different multipartite separability classes.
Here we adopt a recently proposed criterion based on quantum Fisher information (QFI) to identify these fully inseparable seed states~\cite{tms}.
The QFI-based criterion detects multipartite entanglement by systematically optimizing a local operator $\hat{A}$ and testing for violation of the inequality between the QFI and the variances as $F_{Q}(\hat{\rho}_k,\hat{A})\leq4V(\hat{\rho}_k,\hat{A})$, which holds for all $k$-separable states. However, applying this criterion to the detection of multipartite non-Gaussian entanglement fundamentally requires complete knowledge of the density matrix, rendering it infeasible for direct implementation in practical experiments. Nevertheless, we can apply it to the theoretically simulated density matrices in the training set to generate reliable entanglement labels.
In the tripartite case, we randomly generate a dataset of $30\,000$ states, with $10\,000$ states for each entanglement class, including the fully separable class.
\par

\subsection{Correlation patterns from homodyne detection}
We aim to map certain experimentally feasible correlation features, e.g., patterns extracted from homodyne detection, to multipartite entanglement structures described by different multi-partitions $\mathcal{S}$.
Homodyne detection is a projective measurement on eigenstates of orthogonal quadrature operators $\hat x_l=(\hat a_l^\dagger+\hat a_l)$ and $\hat p_l=\text{i}(\hat a_l^\dagger-\hat a_l)$, where $\hat a_l$ is the photon annihilation operator of mode $l\in\{A, B, C\}$~\cite{PhysRevA.47.642,lvovsky2009continuous}. 
According to Born's rule, with $\hat x_l|X_l\rangle=X_l|X_l\rangle$, the probability of obtaining outcome $(X_A,X_B,X_C)$ when simultaneously measuring $\hat x_l$ on the three modes of state $\hat\rho$ is given by $\mathcal{P}(X_A,X_B,X_C)\equiv\langle X_A;X_B;X_C|\hat\rho|X_A;X_B;X_C\rangle$. 
To ensure correlation features are not overlooked when only measuring reduced states, we use a balanced beam splitter $\hat {\mathcal{U}}_{ij}=\text{e}^{\frac{\pi}{4}(\hat a_i^\dagger\hat a_j-\hat a_i\hat a_j^\dagger)}$ to mix modes $i$ and $j$, producing two output modes $M_{ij}$ and $N_{ij}$. Measuring $M_{ij}$ and the remaining unmixed mode $k$ yields two-variable quadrature statistics, thereby extracting more pattern information with fewer measurements.
As an example, by mixing modes $B$ and $C$ and then measuring quadrature $X_A$ of mode $A$ and quadrature $X_{M_{BC}}$ of the mixed mode $M_{BC}$, the joint probability is given by
\begin{align}\label{eq:distribution}
\mathcal{P}(X_A,X_{M_{BC}})&=\langle X_A;X_B\lvert\hat{\mathcal{U}}_{BC}\hat\rho~\hat{\mathcal{U}}_{BC}^\dagger\rvert X_A;X_B\rangle\nonumber\\
&=\langle X_A;\frac{X_B+X_C}{\sqrt{2}}\lvert \hat\rho \rvert X_A;\frac{X_B+X_C}{\sqrt{2}}\rangle.
\end{align} 
Apart from $\mathcal{P}(X_A,X_{M_{BC}})$, we simulate the other three joint probability distributions of $\mathcal{P}(X_A,P_{M_{BC}})$, $\mathcal{P}(P_A,X_{M_{BC}})$, and $\mathcal{P}(P_A,P_{M_{BC}})$ to capture correlation features between subsystems $A$ and $BC$, which corresponds to one of the bipartitions we aim to detect (Supplementary Fig.~5). 
Considering the other two cases with respect to $B-AC$ and $C-AB$ splittings, we simulate a total of twelve joint probability distributions for each state. To process them through the network, we bin every joint probability distribution into a $24\times24$ pixel grid, named a correlation pattern. Although this discretization unavoidably causes some loss of information, the neural network exhibits strong robustness against such degradation, as further discussed in the Discussion. In addition, to account for statistical fluctuations inherent in real experiments, we evaluate the network on datasets simulated from finite sampling points using the Monte Carlo algorithm, which is also given in the Discussion.

\subsection{Network input and training strategy}
For each tripartite state $\hat\rho$, we compute twelve correlation patterns, organized as three groups $\mathcal{C}=\{\mathcal{C}_{AM_{BC}}, \mathcal{C}_{M_{AC}B},\mathcal{C}_{M_{AB}C}\}$. Each group contains four distinct patterns derived from homodyne measurements. Importantly, these patterns, which represent binned joint probability distributions of quadrature measurements (see Eq.~\ref{eq:distribution}) rather than mere second-order correlations, serve as the input to the network. The corresponding ground‑truth multi-partitions $\mathcal{S}_{\text{True}}$, rigorously determined from the full density matrix of each state, serve as the output labels during supervised training.

The network architecture, illustrated in Fig.~\ref{fig:network}(b), is composed of three parallel convolutional sub-networks~\cite{yamashitaConvolutional2018}. Each sub-network independently processes one group of four correlation patterns and condenses them into a feature vector. The three vectors are then combined through fully connected layers, and classified to a final outcome, i.e., the predicted multi-partition class $\mathcal{S}_{\text{Pred}}$. We randomly hold $80\%$ of the $30\,000$ simulated samples as an original training dataset, with the loss function being categorical cross entropy (Supplementary Fig.~11).

During the supervised training phase, the network learns a complex, non-linear mapping from experimentally accessible quadrature data to the entanglement properties defined by the full density matrix. Once trained, it can accurately infer non-Gaussian entanglement directly from such simple inputs in the testing phase. This design bypasses the resource-intensive full state tomography traditionally required for certifying non-Gaussian entanglement, and captures entanglement structures that remain inaccessible to criteria based purely on second-order moments.

\subsection{Quantum data augmentation}
The performance of neural networks heavily depends on the size and diversity of training data. However, for a multipartite non-Gaussian system, the data complexity increases with the number of modes $m$ and the stellar rank $r$~\cite{Chabaud2023PRL}, demanding larger datasets to capture this growing diversity. 
Therefore, efficiently producing comprehensive training data for multipartite CV systems is central to addressing these machine learning problems.
To solve this, we adopt quantum data augmentation, a technique originally developed in classical tasks to synthesize samples from an existing dataset, thereby efficiently enhancing model performance. 
\par
The augmentation operations typically involve subtle variations that do not affect the model’s predictions.
An example of classical data augmentation (DA) via image flipping is shown in Fig.~\ref{fig:aug}(a), where a cat flipped horizontally remains a cat, preserving its label in an image classification task~\cite{shortenSurvey2019}.
Inspired by the success of DA in classical tasks, we introduce a QDA method and explore how augmented data can be constructed for learning quantum systems. 
Similar to classical tasks, these augmentation operations preserve the labels of quantum states.

\begin{figure*}[t]
    \centering
    \includegraphics[width=0.95\linewidth]{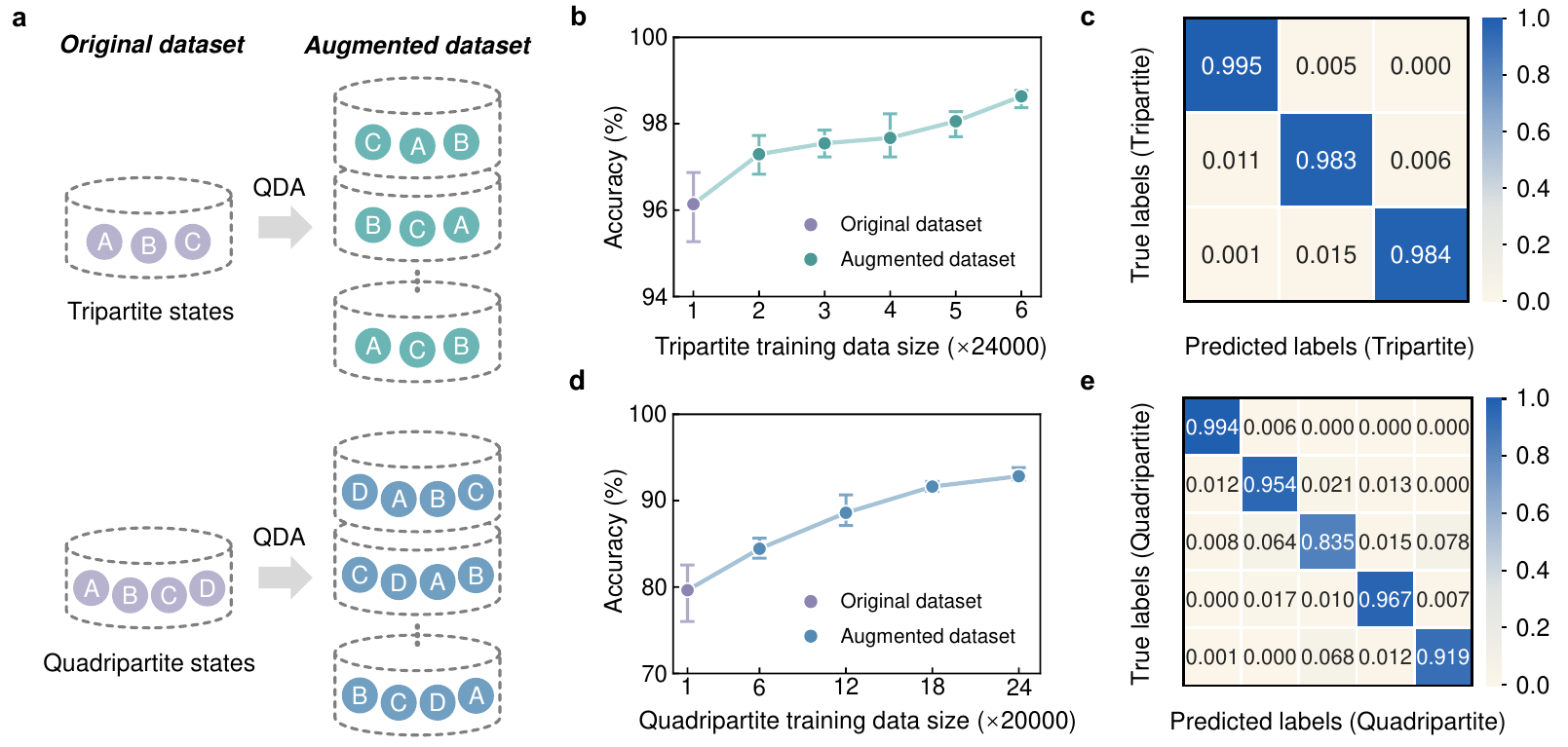}
    \caption{\textbf{Multipartite entanglement structure classification performance after applying QDA.} \textbf{a,} Schematic of QDA for tripartite (top) and quadripartite (bottom) inputs. \textbf{b,} Testing accuracies for tripartite entanglement structure prediction using neural networks trained on the original dataset ($24\,000$ samples, purple marker) and on augmented datasets (green markers). Data are presented as mean values over $n=4$ independent training runs; error bars indicate the minima and maxima across runs. \textbf{c,} Confusion matrix for the tripartite case after QDA with $6\times$ data enlargement. Diagonal entries (top-left to bottom-right) correspond to the prediction accuracies for tripartite partition classes: $1\otimes1\otimes1$, $1\otimes2$, and fully inseparable. \textbf{d,} Testing accuracies for quadripartite entanglement structure prediction using neural networks trained on the original dataset ($20\,000$ samples, purple marker) and on augmented datasets (blue markers). Data are presented as mean values over $n=4$ independent training runs; error bars indicate the minima and maxima across runs. \textbf{e,} Confusion matrix for the quadripartite case after QDA with $24\times$ data enlargement. Diagonal entries represent the accuracies for five partition classes: $1\otimes1\otimes1\otimes1$, $1\otimes1\otimes2$, $2\otimes2$, $1\otimes3$, and fully inseparable.}
    \label{fig:acc}
\end{figure*}

The first QDA operation is based on mode permutation, denoted as $T$. 
Since the entanglement structure is invariant with respect to the mode indexing, the label $\mathcal{S}$ remains unchanged when rearranging the modes of state $\hat\rho$. 
An example of this operation is given in Fig.~\ref{fig:aug}(b). 
The mode permutation $T_{A\leftrightarrow B}$ can be easily achieved by flipping the pattern diagonally and then exchanging correlation patterns in the first two rows and the middle two columns. 
The top panel of Fig.~\ref{fig:aug}(d) shows the simulated correlation patterns $\mathcal{C}^{(1)}$ of a biseparable state $\hat{\rho}^{(1)}_{A|BC}$.  
The middle panel shows the QDA-transformed patterns \(T_{A\leftrightarrow B}(\mathcal{C}^{(1)})\), obtained by applying the mode-permutation rule in Fig.~\ref{fig:aug}(b) directly to the correlation patterns. This transformation preserves the multi-partition label $\mathcal{S}=1\otimes2$, while producing a distinct training sample for the network. 
As a reference, the bottom panel shows the patterns obtained by explicitly permuting modes $A$ and $B$ at the density-matrix level and then calculating the corresponding patterns. 
The two routes produce the same transformed patterns, but QDA avoids the costly density-matrix recalculation quantified in Supplementary Table~1 and obtains them through a simple rearrangement, thereby substantially reducing the sample-generation time.
In principle, for an $m$-mode quantum system, QDA based on mode permutation can expand the size of the dataset to $m!$ times its original size.
\par

Apart from this, some quantum properties can also be used to define QDA operations.
For instance, in our task, a convex combination of any separable states remains separable~\cite{GUHNE20091}, thereby preserving the fully separable partition label $\mathcal{S}=1\otimes1\otimes1$, as shown in Fig.~\ref{fig:aug}(e). 
Meanwhile, correlation patterns of the convexly combined states can also be efficiently obtained, due to the linearity of the expectation value function.
This QDA operation further enlarges our training dataset, and is also valuable for related unsupervised anomaly detection tasks, where separable samples constitute the normal class and entangled samples are treated as anomalies.
\par

\subsection{Entanglement structure classification}
By testing the neural networks on the remaining $20\%$ unseen simulated samples, we assess the classification performance of models trained with the original dataset and augmented datasets, respectively. 
The accuracy measures the ratio of correctly predicted samples to the total number of samples in the test dataset, that is, $N(\mathcal{S}_\text{Pred}=\mathcal{S}_{\text{True}})/N_{\text{total}}$. As a baseline, we apply the QFI-based method to the same homodyne data by first reconstructing the reduced density matrix via maximum-likelihood estimation. Due to the limited correlation patterns available for full state reconstruction, this procedure suffers from significant errors, yielding a classification accuracy of only $0.370$ for the tripartite case. 
We also implement a nearest-neighbor baseline on the input features: for each test correlation pattern, we assign the label of the closest training sample based on Euclidean distance. 
This non-machine-learning baseline achieves accuracies of $0.575$ for tripartite states and $0.485$ for quadripartite states.
\par
By contrast, the neural networks demonstrate strong advantages. When trained on the original dataset only, the network achieves an accuracy of $0.961$ for the tripartite case (purple markers in Fig.~\ref{fig:acc}(b)).
After using augmented data in multiples of the original dataset size, the green markers show significant improvements in the accuracy of entanglement structure prediction. 
It can be raised to $0.986$ when the size of the augmented dataset expands to six times the original size.
A confusion matrix is also evaluated to demonstrate the ability of the network to distinguish among different multi-partitions, as depicted in Fig.~\ref{fig:acc}(c). 
In this matrix, each row represents the samples of a true class and sums to 1, while each column represents samples of a predicted class. 
Using our data-augmented network, the entanglement structure in terms of different multi-partitions can be detected with a prediction accuracy higher than $0.983$ based on limited homodyne measurement data.
\par
It is worth noting that the results presented above are obtained using a combination of both convex combination and mode permutation, with mode permutation contributing 97\% of the augmented samples and convex combination the remaining 3\%. As expected, these two operations yield slightly different effects on network performance; an ablation study comparing models trained with only permutation augmentation, only convex-combination augmentation, and both combined is given in Supplementary III.C.
\par
\begin{table*}[t]
    \centering
    \caption{Neural network performance for different tripartite training dataset sizes.}
    \label{varSize}
    \begin{tabular}{c c c c}
        \hline
        \hline
        Original dataset size & Accuracy & Augmented dataset size & Accuracy \\
        \hline
        4\,000  & 0.905 & 24\,000  & 0.957 \\
        8\,000  & 0.910 & 48\,000  & 0.968 \\
        12\,000 & 0.931 & 72\,000  & 0.972 \\
        16\,000 & 0.950 & 96\,000  & 0.983 \\
        20\,000 & 0.958 & 120\,000 & 0.984 \\
        24\,000 & 0.961 & 144\,000 & 0.986 \\
        \hline
        \hline
    \end{tabular}
\end{table*}

The QDA method is further tested in the quadripartite entanglement structure classification task. 
For each quadripartite state, we mix three modes using two beam splitters and then obtain sixteen correlation patterns by measuring the remaining unmixed mode and one of the mixed outputs (Supplementary Fig.~7).
A neural network with an architecture similar to that used in the tripartite case is trained with an original dataset containing $20\,000$ samples. These samples are evenly distributed across five classes of states, each representing a multi-partition class of fully separable partition $1\otimes1\otimes1\otimes1$, tri-separable partition $1\otimes1\otimes2$, biseparable partitions $2\otimes2$ and $1\otimes3$, and fully inseparable states.
As shown in Fig.~\ref{fig:acc}(d), the network trained with the original dataset only reaches an accuracy of around $0.796$ on $5\,000$ unseen test samples.
After applying QDA, the augmented dataset is expanded 24 times, significantly boosting the accuracy to $0.928$, as depicted by the blue markers. The confusion matrix in Fig.~\ref{fig:acc}(e) highlights the QDA improvements, where the augmented network nearly perfectly classifies fully separable states and achieves $0.919$ accuracy in identifying fully quadripartite inseparable states.
These accuracies can be further improved by feeding in additional correlation patterns, especially for the class $\mathcal{S}=2\otimes2$, as detailed in the Discussion.
Evidently, with the enhancement of QDA, the neural network becomes more powerful in classifying multipartite entanglement structures across all possible multi-partitions.
\par

\section{Discussion}
\begin{figure*}[p]
    \centering
    \includegraphics[width=0.95\linewidth]{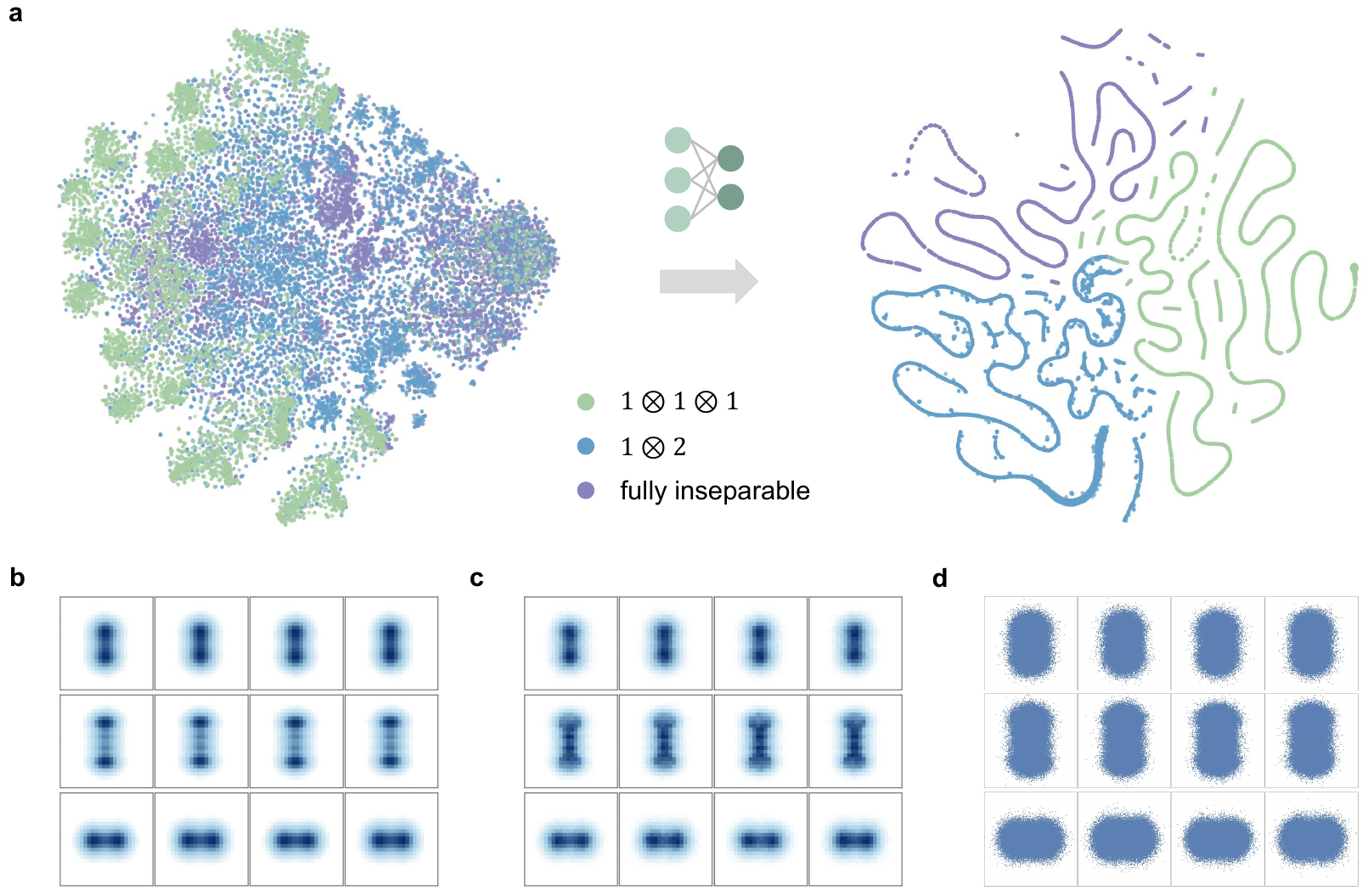}
    \caption{\textbf{Robustness to discretization and statistical fluctuations in correlation patterns.} \textbf{a,} Two-dimensional t-SNE visualization of tripartite correlation patterns before (left) and after (right) processing by the augmented neural network. The t-SNE map of the original discretized correlation patterns shows strong overlap among different multi-partition classes, whereas the corresponding projection of the final-layer network features yields clearly separated clusters, suggesting that discretization-induced information loss does not prevent effective classification. \textbf{b,} Ideal correlation patterns $\mathcal{C}_{\text{ideal}}$ binned from the exact joint probability distribution. \textbf{c,} Noisy correlation patterns $\mathcal{C}_{\text{MC}}$ binned from Monte Carlo sampling. \textbf{d,} Monte Carlo sampling of $10^5$ statistically independent points for each pattern, simulating a set of joint homodyne detection events.}
    \label{fig:tsne}
\end{figure*}
Complementing these results, we now examine the performance and adaptability of our network under practical conditions. In particular, we investigate the influence of original training-set size, the inclusion of additional correlation patterns, the potential information loss due to pattern discretization, and statistical fluctuations from finite measurements. We further assess the network’s generalization ability to typical experimental noise, as well as to unknown infinite‑stellar‑rank non-Gaussian states. Finally, we demonstrate the scalability of our approach to five‑mode CV systems. These analyses collectively highlight the robustness, flexibility, and broader applicability of the data-augmented network for multipartite entanglement classification.
\par

First, we vary the size of the original training set for tripartite states and retrain the network. As illustrated in Table~\ref{varSize}, a key observation is that the largest accuracy gap between the original and augmented datasets occurs at $8\,000$ samples, where the augmented dataset achieves $0.968$, compared to $0.910$ for the original simulated dataset. However, as the original dataset size increases, the gap narrows, and the accuracy after augmentation converges to $0.986$, suggesting that the marginal benefit of augmentation decreases with larger original datasets. This indicates that while data augmentation is beneficial across all dataset sizes, it has a more significant impact when the original dataset is smaller.
\par
Second, note that in Results, each joint probability distribution of quadripartite states is obtained after mixing three modes; hence, only correlation features for the $A-BCD$, $B-ACD$, $C-ABD$, and $D-ABC$ splittings are captured. However, for states with a biseparable partition $2\otimes2$, the separability cannot be perfectly represented by these input data. Therefore, the detection accuracy for states with $\mathcal{S}=2\otimes2$ only reaches $0.835$, as shown in Fig.~\ref{fig:acc}(e). Even when training is restricted to states with finite stellar rank, the detection accuracy for $\mathcal{S} = 2 \otimes 2$ states remains limited to $0.844$. To further improve the accuracy of classifying quadripartite entanglement structures, especially for the class $\mathcal{S}=2\otimes2$, we fed the neural network with new correlation patterns, which are binned from sixteen additional joint probability distributions providing correlation features of $A-BC$, $B-AC$, $C-AB$, and $D-AB$ splittings (Supplementary Fig.~8). 
With these additional distributions, the accuracy for detecting states in the biseparable partition $\mathcal{S}=2\otimes2$ is improved to $0.922$ when the size of the augmented dataset expands to 24 times the original size. 
Correspondingly, the average accuracy for detecting entanglement structures of quadripartite states is improved to $0.969$.
\par

Third, to assess whether the discretization of patterns leads to information loss that could degrade network performance, we apply the t-distributed stochastic neighbor embedding (t-SNE) algorithm to visualize the high-dimensional input correlation patterns. This technique projects each pattern onto a two-dimensional manifold, where the proximity between points reflects the similarity of their underlying correlation structures. 
As shown in Fig.~\ref{fig:tsne}(a), the raw discretized feature space exhibits significant overlap among samples from different entanglement-structure classes. 
This initial clustering indicates that discretization can indeed lead to near-degenerate representations that are difficult to distinguish via direct observation. 
After processing by the trained neural network, however, the same dataset becomes markedly well separated. Previously overlapping samples are mapped into distinct, isolated clusters in the learned feature space. These results demonstrate that, within the regime considered, the information loss inherent to discretization does not impose a fundamental limitation. Rather, the network’s nonlinear feature extraction effectively resolves these near-degeneracies, enabling robust identification of entanglement structures from experimentally accessible correlation patterns.
\par
Finally, we assess the robustness against statistical fluctuations in correlation patterns, mimicking finite-sample homodyne data in actual experiments. 
To investigate their impact, we further simulate the limited measurement outcomes via a Monte Carlo sampling method and find that even with finite sampling points, the data-augmented network can still maintain high accuracy.\par

To be more specific, holding 100 new tripartite samples with statistical fluctuations as the test dataset, we evaluate the neural network's performance. 
Focusing on states with finite stellar rank, we denote input samples derived from ideal correlation patterns as $\mathcal{C}_{\text{ideal}}$, and those incorporating statistical fluctuations as $\mathcal{C}_{\text{MC}}$, depicted in Figs.~\ref{fig:tsne}(b) and (c), respectively. These fluctuations, arising from homodyne detection sampling, are simulated using a Monte Carlo process, as illustrated in Fig.~\ref{fig:tsne}(d).
Training with $24\,000$ ideal samples $\{\mathcal{C}_{\text{ideal}}\}$ yields accuracies of $0.929$ and $0.976$ on the ideal test dataset before and after QDA, respectively. 
However, when tested on the 100 non-ideal samples $\{\mathcal{C}_{\text{MC}}\}$, the same network architecture achieves accuracies of $0.910$ and $0.950$ before and after QDA, respectively. 
Hence, imperfect knowledge of the joint probability distributions only slightly weakens the performance of the neural network. 
In addition, introducing a small number of samples with statistical fluctuations into the training dataset can help the network learn to recognize and account for the impact of fluctuations on entanglement detection. After using $23\,700$ ideal samples $\{\mathcal{C}_{\text{ideal}}\}$ and 300 non-ideal samples $\{\mathcal{C}_{\text{MC}}\}$ for training, we still test the neural network on the 100 non-ideal test samples. 
The network achieves improved accuracies of $0.930$ and $0.970$ before and after data augmentation, respectively. 

To assess generalization ability, we evaluate our trained network under two challenging conditions beyond its training distribution. First, despite being trained only on simulated data with photon loss, the network maintains high robustness when tested on samples incorporating thermal and phase noise, achieving an accuracy of $0.980$ on hundreds of noisy tripartite states (Supplementary I.D). Second, we test the model on $1\,000$ unseen non-Gaussian states generated via spontaneous parametric down-conversion (Supplementary Fig.~3), which is a physically distinct class of infinite-stellar-rank states absent from the training set. 
Without fine-tuning the network or using prior information about the states, the network achieves near-perfect accuracy in identifying their entanglement structures. 
These results demonstrate that our model has learned robust, transferable features of multipartite entanglement that generalize across key experimental imperfections (including photon loss, thermal noise, phase noise, and finite sampling) as well as to states beyond the original training distribution.  

\begin{figure}[t]
    \centering
    \includegraphics[width=0.95\linewidth]{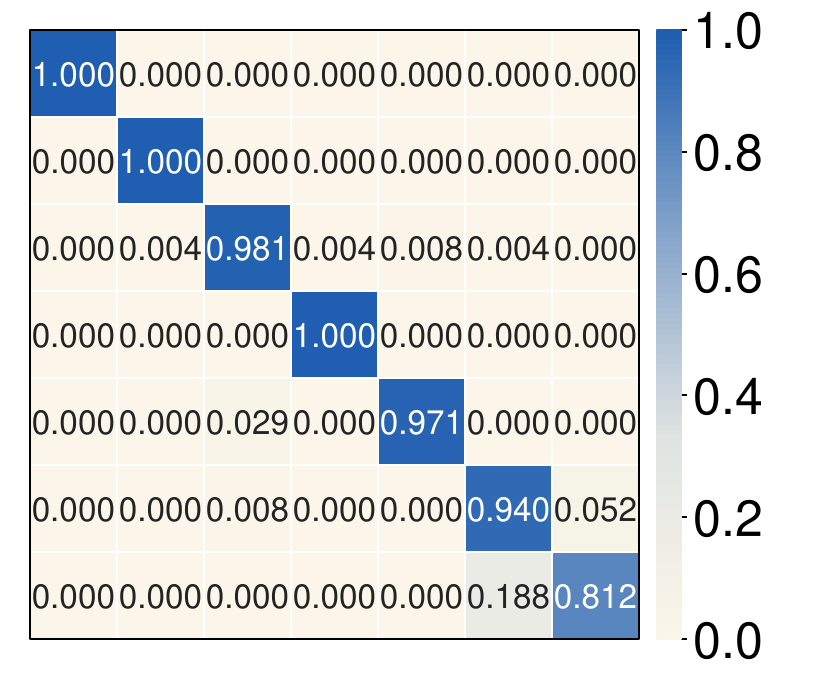}
    \caption{\textbf{Confusion matrix for the five-mode case after QDA with $100$-fold data enlargement.} The diagonal elements represent prediction accuracies for partitions $1\otimes1\otimes1\otimes1\otimes1$, $1\otimes1\otimes1\otimes2$, $1\otimes2\otimes2$, $1\otimes1\otimes3$, $1\otimes4$, $2\otimes3$, and fully inseparable, respectively.}
    \label{fig:5mode}
\end{figure}

Beyond feature engineering, a critical consideration for any quantum characterization framework is its scalability to larger systems. For an $m$-mode system, the complexity of entanglement structures is fundamentally determined by the integer partitions of $m$. 
In the five-mode case, this yields seven distinct entanglement classes. To evaluate scalability, we extend our analysis to $7\,000$ simulated five-mode states ($1\,000$ per class), spanning stellar ranks $r \in \{0, 1, 2, 3, +\infty\}$, with 80\% of the data used for training. The input correlation patterns are generated following an $(m-1)$-mode beam splitter mixing protocol (Supplementary Fig.~9), and we apply a 100-fold augmentation by performing distinct permutations within the training set. The impact of QDA remains substantial as system size increases: classification accuracy for five-mode states improves from 76.7\% to 92.4\%. As illustrated in the confusion matrix (Fig.~\ref{fig:5mode}), several partitions achieve near-perfect identification. Nevertheless, further scaling to larger systems will incur additional computational costs. Generating and labelling diverse CV states, constructing informative features, and training larger networks all become increasingly expensive as the number of modes and non-Gaussian complexity grow. Hence, substantial further scaling will require careful management of the overall computational budget.
\section{Conclusion}
In conclusion, we present a data-augmented deep learning approach for classifying quantum entanglement structures of general multipartite CV states with both finite and infinite stellar ranks, using only feasible homodyne measurement data. This procedure involves training a convolutional neural network, where the network is fed with diverse synthetic datasets generated through the QDA method, rooted in fundamental physical principles. While the present work focuses on the classification of partition‑based entanglement structures, our framework is inherently extensible to finer‑grained discrimination of distinct entanglement types within a given partition. By transitioning from a permutation-invariant mapping to a permutation-equivariant one, the network can effectively distinguish between specific configurations, such as different bipartite splits within the biseparable class (Supplementary III.E).
\par

More broadly, this strategy can be generalized to other complex multipartite scenarios by exploiting the invariance of target quantum properties under specific operations, a concept analogous to ``free operations" in quantum resource theory. Such guidance from resource theory could inform the development of QDA strategies. Future work may fruitfully explore hybrid designs that incorporate this resource invariance, particularly in conjunction with recently developed equivariant neural network architectures~\cite{PRXQuantum.5.020328,PhysRevA.109.022405}. Critically, QDA-enhanced networks not only capture underlying entanglement features but also serve as practically valuable tools for quantum information processing tasks.
\par

\section{Methods}
\subsection{The stellar formalism}
To randomly generate $m$-mode Gaussian and non-Gaussian seed states $\hat\sigma_m$, we first characterize them using the stellar formalism~\cite{Chabaud2020stellar,Chabaud2021classical,Chabaud2022holomorphic}. 
In this formalism, we analyze an $m$-mode normalized pure state $|\psi\rangle$ in terms of its stellar function $F_\psi^{\star}(\mathbf{z}) \equiv \text{e}^{\frac{1}{2}\|\mathbf{z}\|^2}\left\langle \mathbf{z}^*|\psi\right\rangle$, where $|\mathbf{z}\rangle$ is a coherent state with complex amplitude $\mathbf{z}$. 
The stellar rank $r$ of $|\psi\rangle$ is defined as the number of zeros of $F_\psi^{\star}(\mathbf{z})$, representing the minimal non-Gaussian operational cost to engineer the state from the vacuum.
\par
By introducing the notation $\overline{\mathbb{N}} = \mathbb{N} \cup \{+\infty\}$, the stellar hierarchy of continuous variable states is induced by the stellar rank $r\in\overline{\mathbb{N}}$.
For instance, $r=0$ means that the state is Gaussian, while $r=1$ corresponds to a class of non-Gaussian states that contain both single-photon added and subtracted states. 
Another important and experimentally accessible class of non-Gaussian states, cat states, is of infinite stellar rank $r\rightarrow\infty$.
\par

\subsection{Correlation patterns transformation under mode permutation}  
For the tripartite quantum system, when exchanging modes $A$ and $B$, the joint probabilities transform as follows

\begin{equation}\label{eq:tri-trans}
\begin{array}{ccc}
     \langle X_A;\frac{X_B+X_C}{\sqrt{2}}| \hat\rho | X_A;\frac{X_B+X_C}{\sqrt{2}}\rangle\\ 
     \langle \frac{X_A+X_C}{\sqrt{2}};X_B| \hat\rho | \frac{X_A+X_C}{\sqrt{2}};X_B\rangle\\  
     \langle \frac{X_A+X_B}{\sqrt{2}};X_C| \hat\rho | \frac{X_A+X_B}{\sqrt{2}};X_C\rangle\\[4pt]
     \downarrow{A\leftrightarrow B}\\[4pt]
      \langle X_B;\frac{X_A+X_C}{\sqrt{2}}| \hat\rho | X_B;\frac{X_A+X_C}{\sqrt{2}}\rangle\\
       \langle \frac{X_B+X_C}{\sqrt{2}};X_A| \hat\rho | \frac{X_B+X_C}{\sqrt{2}};X_A\rangle\\
     \langle \frac{X_B+X_A}{\sqrt{2}};X_C| \hat\rho | \frac{X_B+X_A}{\sqrt{2}};X_C\rangle. \\
\end{array}
\end{equation}
This transformation can be efficiently realized by swapping the first two original probability distributions and transposing them, respectively, leaving the remaining probability distribution unchanged. 
\par

For the quadripartite quantum system, when exchanging modes $A$ and $B$, the joint probabilities will be transformed as below

\begin{align}
\begin{array}{ccc}
     \langle X_A;\frac{X_B+X_C+X_D}{\sqrt{3}}| \hat\rho| X_A;\frac{X_B+X_C+X_D}{\sqrt{3}}\rangle \\
     \langle \frac{X_A+X_C+X_D}{\sqrt{3}};X_B| \hat\rho | \frac{X_A+X_C+X_D}{\sqrt{3}};X_B\rangle \\
     \langle \frac{X_A+X_B+X_D}{\sqrt{3}};X_C| \hat\rho | \frac{X_A+X_B+X_D}{\sqrt{3}};X_C\rangle \\
     \langle \frac{X_A+X_B+X_C}{\sqrt{3}};X_D| \hat\rho | \frac{X_A+X_B+X_C}{\sqrt{3}};X_D\rangle \\[4pt]
     \downarrow{A\leftrightarrow B}\\[4pt]
     \langle X_B;\frac{X_A+X_C+X_D}{\sqrt{3}}| \hat\rho| X_B;\frac{X_A+X_C+X_D}{\sqrt{3}}\rangle\\
     \langle \frac{X_B+X_C+X_D}{\sqrt{3}};X_A| \hat\rho | \frac{X_B+X_C+X_D}{\sqrt{3}};X_A\rangle\\
     \langle \frac{X_B+X_A+X_D}{\sqrt{3}};X_C| \hat\rho | \frac{X_B+X_A+X_D}{\sqrt{3}};X_C\rangle \\
     \langle \frac{X_B+X_A+X_C}{\sqrt{3}};X_D| \hat\rho | \frac{X_B+X_A+X_C}{\sqrt{3}};X_D\rangle.\\
\end{array}
\end{align}

This transformation can also be efficiently realized by swapping the first two original probability distributions and transposing them, respectively, while leaving the other two probability distributions unchanged. \par

For the additional quadripartite joint probability distributions, the transformation is similar to that in Eq.~\ref{eq:tri-trans}. The same transformation can be applied to the correlation patterns, which are binned from these joint probability distributions. 
Correspondingly, the correlation patterns are also flipped and exchanged according to the above transformations.
Although index swapping does not change the characterization of the quantum states, the rearranged correlation patterns are treated as new samples by the neural network.
Therefore, the transformation results in a modified sample that can still be assigned to the original entanglement structure label $\mathcal{S}$.
\par

\textbf{Data availability:}
The datasets generated and analyzed in this study are available in the Code Ocean capsule associated with this article at \url{https://doi.org/10.24433/CO.9492084.v2} (ref.~\cite{data_code}).

\textbf{Code availability:}
The code used for quantum data augmentation, neural-network training, and model evaluation is available in the Code Ocean capsule at \url{https://doi.org/10.24433/CO.9492084.v2} (ref.~\cite{data_code}). The corresponding GitHub repository has been archived on Zenodo at \url{https://doi.org/10.5281/zenodo.20928779} (ref.~\cite{github_code}). The neural networks were implemented using TensorFlow~\cite{TensorFlow}.

\textbf{Funding}\par
This work was supported by Beijing Natural Science Foundation (Grant No.~Z240007), National Natural Science Foundation of China (No.~12125402, No.~12534016, No.~62575232, No.~12350006, No.~12405022, No.~12547180, and No.~12474256), and Quantum Science and Technology-National Science and Technology Major Project (Grant No.~2024ZD0302401, No.~2021ZD0301500, and No.~2025ZD0301000).

\textbf{Acknowledgements}\par
The authors have no additional acknowledgements.

\textbf{Author contributions}\par
X.G., Y.X., and Q.H. conceived the original idea. Q.H. managed the project. X.G. performed the numerical simulations and developed the neural network model. M.T. proposed the optimized QFI criterion algorithm to label the quantum states. All the authors discussed the results and contributed to the final version of the paper.

\textbf{Competing interests}\par
The authors declare no competing interests.

\textbf{Additional information}\par
\textbf{Supplementary information} The online version contains supplementary material available at xxx.

\textbf{Correspondence and requests for materials} should be addressed to Yu~Xiang or Qiongyi~He.

\bibliographystyle{sn-nature}

\clearpage
\onecolumn

\setcounter{section}{0}
\setcounter{subsection}{0}

\renewcommand{\thesection}{\Roman{section}}
\renewcommand{\thesubsection}{\Alph{subsection}}

\makeatletter
\renewcommand{\@seccntformat}[1]{%
  \csname the#1\endcsname.\quad}
\makeatother

\addtocontents{toc}{\protect\setcounter{tocdepth}{2}}

\begin{center}
    {\LARGE\bfseries Supplementary Information}
\end{center}

\vspace{0.5cm}

\tableofcontents
\clearpage

\section{Generation process of quantum states}

\subsection{Seed states with finite stellar rank}\label{stateGeneration} 
Any multimode pure state $|\psi\rangle$ with finite stellar rank $r$ can be decomposed into $|\psi\rangle=\hat G|C\rangle$, where $\hat G$ is a Gaussian operator acting on the state $|C\rangle$, which is called a core state; it is a normalized pure quantum state with multivariate polynomial stellar function of degree $r$, equal to the stellar rank of $|C\rangle$. 
For example, the stellar function of a three-mode core state $|012\rangle$ is given by $\frac{1}{\sqrt{2}}z_2z_3^2$, whose degree is equal to the stellar rank $r=3$ of the state.
It then follows immediately that Gaussian operations $\hat G$ must preserve the stellar rank~\cite{Chabaud2021classical}. \par

\begin{figure}[bht]
    \centering
    \includegraphics[width=0.85\linewidth]{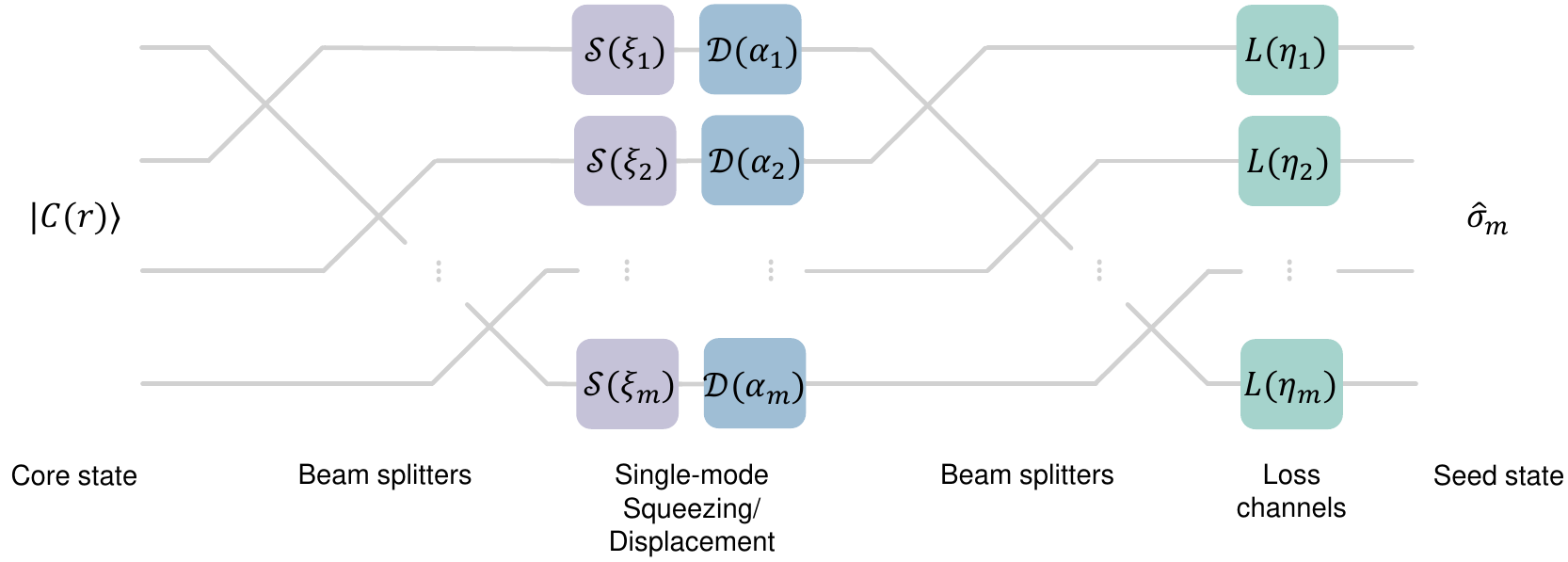}
    \caption{The generation process of an $m$-mode seed state $\hat\sigma_m$.}
    \label{fig:seed}
\end{figure}
Our randomly generated seed states follow the above decomposition, which begins with a core state $|C(r)\rangle$ with a given stellar rank $r$ and random complex superposition coefficients in the Fock basis. 
We restrict the stellar rank to $r\leq3$, which includes the most common Gaussian and non-Gaussian states in experiments. 
According to the Williamson decomposition and the Bloch-Messiah decomposition, an $m$-mode Gaussian unitary operation $\hat G$ can be decomposed as $\hat G=\mathcal{\hat U}(\varphi)\left( \prod\limits_{l=1}^m \mathcal{\hat S}_l(\xi_l)\mathcal{\hat D}_l(\alpha_l)\right)\mathcal{\hat V}(\phi)$, where $\mathcal{\hat S}_l(\xi_l)=\text{e}^{\frac{1}{2}(\xi_l^*\hat a_l^2-\xi_l\hat a_l^{\dagger2})}$ is a single-mode squeezing operator with complex squeezing parameter $\xi_l$ acting on mode $l$, and $\mathcal{\hat D}_l(\alpha_l)=\text{e}^{\alpha_l\hat a^\dagger_l-\alpha_l^*\hat a_l}$ is a displacement operator with complex displacement amplitude $\alpha_l$ acting on mode $l$. 
$\mathcal{\hat U}(\varphi)$ and $\mathcal{\hat V}(\phi)$ are passive linear optical transformations over $m$ modes, consisting of $(m-1)$ beam splitters with complex coupling coefficients $\varphi$ and $\phi$, respectively.\par

Losses are also added to each mode of the pure state
$|\psi\rangle=\hat G|C(r)\rangle$, using a single-mode loss channel $\hat L_l(\eta_l) = \sqrt{\frac{(1-\eta_l)^n}{n! \eta_l^n}} \hat{a}_l^n e^{\frac{1}{2}  \hat{a}_l^\dagger \hat{a}_l\text{ln}\eta_l}$ as described in Refs.~\cite{Eaton2022measurementbased,LossChannelLect}, representing an $n$-photon loss channel with efficiency $\eta_l$. Here we truncate the degree at $n=10$.
The entire quantum circuit is shown in Fig.~\ref{fig:seed}, generating an $m$-mode seed state $\hat\sigma_m$ with stellar rank $r$ and several randomly selected free parameters, $\mathbf{\varphi}$, $\mathbf{\xi}$, $\mathbf{\alpha}$, and $\mathbf{\phi}$, given by
\begin{align}\label{eq:state}
\hat\sigma_m=\left(\prod_{l=1}^m\hat L_l(\eta_l)\right)\hat G |C(r)\rangle\langle C(r)|\hat G^\dagger \left(\prod_{l=1}^m\hat L_l^\dagger(\eta_l)\right).
\end{align}\par

\subsection{Cat states with infinite stellar rank}
In addition to multipartite quantum states with finite stellar rank, non-Gaussian states with infinite stellar rank, such as entangled cat states, are also considered. These states can be detected by modular variables~\cite{ModularOperator,PhysRevLett.106.210501}. Below, we demonstrate the state generation process of multipartite states with infinite stellar rank, where multimode cat states and single-mode coherent states are employed as seed states.\par

\begin{figure}[H]
    \centering
    \includegraphics[width=0.85\linewidth]{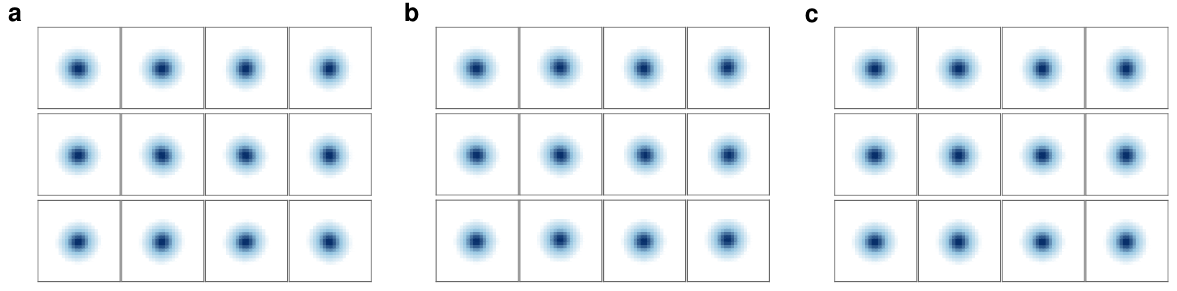}
    \caption{\textbf{Correlation patterns of lossy tripartite cat states.} \textbf{a,} fully entangled cat state; \textbf{b,} biseparable cat states, and \textbf{c,} fully separable cat state with the same parameters $|\alpha|=0.231$, $|\beta|=0.164$,$|\gamma|=0.159$, $\phi_1=6.148$, $\phi_2=4.362$, $\phi_3=1.436$ and efficiency coefficients $\eta_1=0.926,\eta_2=0.865,\eta_3=0.707$ of the loss channels for each mode.}
    \label{fig:catState}
\end{figure}

For the tripartite case, we randomly generate $5\,000$ pure fully inseparable cat states $|\psi\rangle_{\text{fi}}=1/\sqrt{N_{\text{fi}}}(|\alpha\rangle|\beta\rangle|\gamma\rangle+|-\alpha\rangle|-\beta\rangle|-\gamma\rangle)$. Losses are added to these states with the same loss channel as in Sec.~\ref{stateGeneration}, and the lossy entangled states are identified by the QFI entanglement criterion. In addition, we also randomly generate $5\,000$ lossy biseparable cat states of the form $|\psi\rangle_{\text{bi}}=1/\sqrt{N_{\text{bi}}}(|\alpha\rangle|\beta\rangle+|-\alpha\rangle|-\beta\rangle)|\gamma\rangle$ and $5\,000$ lossy fully separable cat states $|\psi\rangle_{\text{fs}}=1/\sqrt{N_{\text{fs}}}(|\alpha\rangle+|-\alpha\rangle)(|\beta\rangle+|-\beta\rangle)(|\gamma\rangle+|-\gamma\rangle)$ to form a simulated infinite-stellar-rank dataset. Here, $|\alpha\rangle$, $|\beta\rangle$, and $|\gamma\rangle$ are coherent states with complex amplitude $\alpha=|\alpha|\text{e}^{\text{i}\phi_1}$, $\beta=|\beta|\text{e}^{\text{i}\phi_2}$, and $\gamma=|\gamma|\text{e}^{\text{i}\phi_3}$. The parameter range is set to $|\alpha|\in[0,0.4]$, $|\beta|\in[0,0.4]$, $|\gamma|\in[0,0.4]$, and $\phi_l\in[0,2\pi)$, with $l=1,2,3$. $N_{\text{fi}}$, $N_{\text{bi}}$ and $N_{\text{fs}}$ are the normalization constant. Correlation patterns and entanglement structure label pairs $\{\mathcal{C},\mathcal{S}\}$ are then obtained; $80\%$ are fed into the neural network for training and $20\%$ for testing. Fig.~\ref{fig:catState} shows the correlation patterns of these three types of states. \par


For the quadripartite case, we consider five distinct entanglement structures and randomly generate a total of $12\,500$ pure states, with $2\,500$ samples for each structure. Among them, we include fully inseparable cat states of the form $|\psi\rangle_{\text{fi}}=1/\sqrt{N_{\text{fi}}}(|\alpha\rangle|\beta\rangle|\gamma\rangle|\delta\rangle+|-\alpha\rangle|-\beta\rangle|-\gamma\rangle|-\delta\rangle)$, $1\otimes3$-partitioned states $|\psi\rangle_{1|3}=1/\sqrt{N_{1|3}}(|\alpha\rangle|\beta\rangle|\gamma\rangle+|-\alpha\rangle|-\beta\rangle|-\gamma\rangle)|\delta\rangle$, $2\otimes2$-partitioned states $|\psi\rangle_{2|2}=1/\sqrt{N_{2|2}}(|\alpha\rangle|\beta\rangle+|-\alpha\rangle|-\beta\rangle)(|\gamma\rangle|\delta\rangle+|-\gamma\rangle|-\delta\rangle)$, $1\otimes1\otimes2$-partitioned states $|\psi\rangle_{1|2|1}=1/\sqrt{N_{1|2|1}}(|\alpha\rangle|\beta\rangle+|-\alpha\rangle|-\beta\rangle)|\gamma\rangle|\delta\rangle$, and fully separable states $|\psi\rangle_{\text{fs}}=1/\sqrt{N_{\text{fs}}}(|\alpha\rangle+|-\alpha\rangle)(|\beta\rangle+|-\beta\rangle)(|\gamma\rangle+|-\gamma\rangle)(|\delta\rangle+|-\delta\rangle)$. The parameter ranges of coherent states $|\alpha\rangle$, $|\beta\rangle$, $|\gamma\rangle$, and $|\delta\rangle$ are set to be the same as in the tripartite case. $N_{\text{fi}}$, $N_{1|3}$, $N_{2|2}$, $N_{1|2|1}$, and $N_{\text{fs}}$ are the normalized parameters. 

\subsection{\red{Testing with states generated via spontaneous parametric down-conversion (SPDC) process}}
\red{To further assess the generalization capability of our trained neural network, we perform additional testing on a physically prominent class of non-Gaussian states generated via the spontaneous parametric down-conversion (SPDC) process. These SPDC-generated samples also possess an infinite stellar rank and represent a state distribution that the network has not previously encountered.}

\red{We model the three-mode system using the following SPDC Hamiltonian:}
\begin{align}
    \red{\hat{H} = \chi_1 \hat{a} \hat{b}^2 + \chi_2 \hat{b} \hat{c}^2 + \chi_3 \hat{c} \hat{a}^2 + \text{H.c.},}
\end{align}
\red{where $\hat{a}$, $\hat{b}$, and $\hat{c}$ are the bosonic annihilation operators for the respective modes. The parameters $\chi_1, \chi_2$, and $\chi_3$ are randomized within the range of $[0.01, 0.04]$ to produce a diverse evaluation set of $1,000$ unknown non-Gaussian samples.}

\red{Remarkably, despite being trained only on a mixture of finite-rank states and cat states, our neural network achieves near-perfect accuracy in identifying the entanglement structures of these SPDC samples. This performance is attained without any fine-tuning of the network or prior information about the state preparation. These results suggest that the network has successfully learned robust, intrinsic features of multipartite entanglement, offering a high degree of generalization for the complex non-Gaussian states currently accessible in experimental settings.}
\begin{figure}
    \centering
    \includegraphics[width=0.3\linewidth]{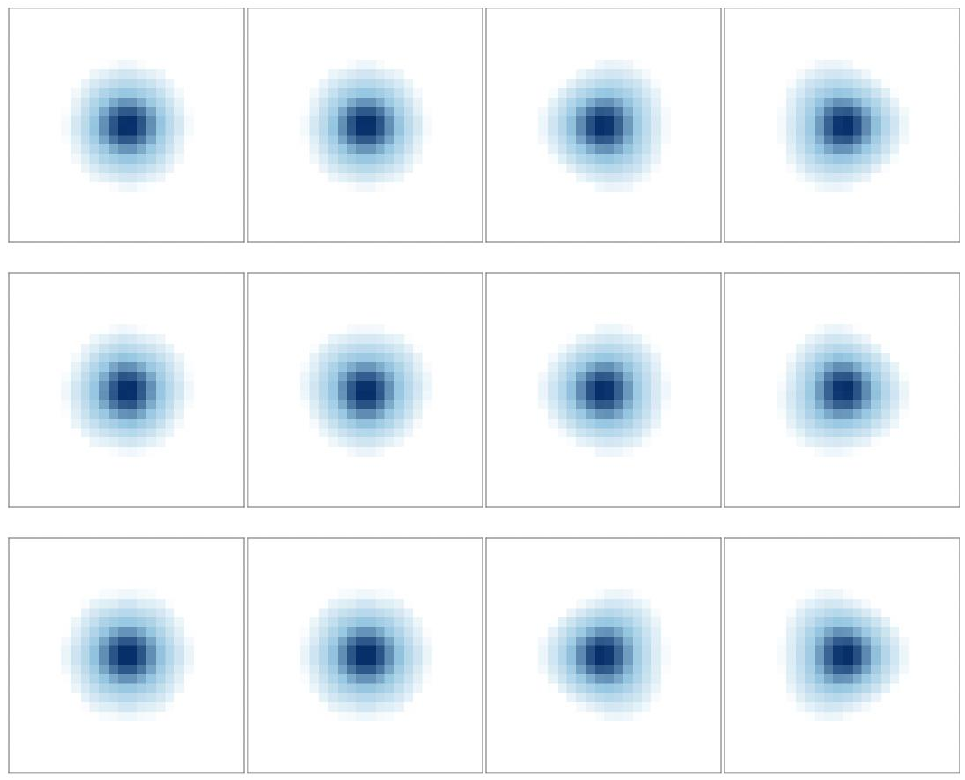}
    \caption{\red{Correlation patterns of states via the SPDC process.}}
    \label{fig:SPDC}
\end{figure}
\subsection{\red{Testing with thermal noise and phase noise}}
\red{To evaluate the performance of our neural network under realistic experimental conditions, we extend our simulations beyond photon loss in Eq.~\ref{eq:state} to explicitly incorporate thermal noise and phase noise, which are primary noise sources in continuous-variable (CV) systems.
Thermal noise is modeled by the following quantum map}
\begin{align}
    \red{\hat{\rho} \mapsto \mathcal{E}(\hat{\rho}) = \sum_{k=0}^{\infty} B_k \hat{\rho} B_k^{\dagger},}
\end{align} 
\red{where $B_k(G) = \sqrt{\frac{1}{k!G} \left( \frac{G-1}{G} \right)^k} \hat{a}^{\dagger k} G^{-\hat{n}/2}$ are the Kraus operators of a quantum-limited amplifier. }
\par
\red{Additionally, phase noise is incorporated through the map}
\begin{align}
    \red{\hat{\rho} \mapsto \mathcal{E}(\hat{\rho}) = \int_{\mathbb{R}} d\varphi f(\varphi) \hat{U}(\varphi) \hat{\rho} \hat{U}^{\dagger}(\varphi),}
\end{align}
\red{where $\hat{U}(\varphi) = \exp(-i\varphi \hat{a}^{\dagger} \hat{a})$ is the phase-shift operator and $f(\varphi)$ represents a weight function describing the noise distribution. 
In our dataset, both uniform and Gaussian distribution models are considered to match common experimental settings.
We test the network on a dataset consisting of hundreds of samples subjected to these challenging noisy channels. 
It is worth emphasizing that the neural network employed here is trained on the augmented tripartite dataset described in the main text, which incorporates only photon loss.
Remarkably, the neural network retains high robustness, maintaining a classification accuracy of $98.0\%$ on a test set of hundreds of noisy tripartite states. 
This result confirms that the features learned by the network from discretized homodyne data are sufficiently robust to generalize to practical CV experimental scenarios where multiple noise sources coexist.}
\section{Correlation patterns generation}
\subsection{Beam splitter transformation}
A beam splitter operator spanning on modes $i$ and $j$ is written as $\hat {\mathcal{U}}_{ij}(\theta)=\text{e}^{\theta(\hat a_i^\dagger\hat a_j-\hat a_i\hat a_j^\dagger)}$, where $t=\text{cos}\theta$ is the transmission coefficient in amplitude and thus the square of coefficient in energy. 
\begin{align}\label{eq:anni}
    \hat {\mathcal{U}}_{ij}^\dagger(\theta)\left(\begin{array}{cc}\hat a_i\\\hat a_j \end{array}\right)\hat {\mathcal{U}}_{ij}(\theta)=
    \left(\begin{array}{cc}\hat a_i\text{cos}\theta+\hat a_j\text{sin}\theta\\-\hat a_i\text{sin}\theta+\hat a_j\text{cos}\theta \end{array}\right)=
    \mathcal{T}(\theta)\left(\begin{array}{cc}\hat a_i\\\hat a_j \end{array}\right).
\end{align}
where $\mathcal{T}(\theta)=\left(\begin{array}{cc}\text{cos}\theta & \text{sin}\theta \\-\text{sin}\theta & \text{cos}\theta\end{array}\right)$ is a transmission matrix. To simplify, we denote ${\mathcal{U}}_{ij}(\theta)$ as ${\mathcal{U}}$ and $\mathcal{T}(\theta)$ as $\mathcal{T}$. Note that the transmission matrix is unitary; hence, $\mathcal{T}^\dagger=\mathcal{T}^{-1}$. \par

Consider an input two-mode density matrix $\hat\rho_{\text{in}}[\hat{\boldsymbol{a}},\hat {\boldsymbol{a}}^\dagger]$ being regarded as a function of operators $\hat{\boldsymbol{a}}=(\hat a_1,\hat a_2)^T$, then the output density matrix after the beam splitter transformation is given by 
\begin{align}
\hat\rho_{\text{out}}[\hat{\boldsymbol{a}}]=\hat {\mathcal{U}}\hat\rho_{\text{in}}[\hat{\boldsymbol{a}}]\hat {\mathcal{U}}^\dagger
=\hat\rho_{\text{in}}[\hat {\mathcal{U}}\hat{\boldsymbol{a}}\hat {\mathcal{U}}^\dagger]
=\hat\rho_{\text{in}}[{\mathcal{T}}^\dagger\hat{\boldsymbol{a}}], 
\end{align}
which is the inverse transformation of annihilation operator transformation in Eq.~\ref{eq:anni}~\cite{agarwal2012quantum}.\par

Defining a vector $\boldsymbol{\alpha}=[\alpha_1,\alpha_2]^T$ in phase space, the Wigner function of a two-mode state is written as $W(\alpha)=\text{Tr}[\hat\rho\delta(\hat{\boldsymbol{a}},\boldsymbol{\alpha})]$. If we denote the Wigner function of input state $\hat\rho_{\text{in}}$ by $W_{\text{in}}(\boldsymbol{\alpha})$ and the Wigner function of output state $\hat\rho_{\text{out}}$ by $W_{\text{out}}(\boldsymbol{\alpha})$, then 
\begin{equation}
\begin{aligned}\label{eq:wigner}
    W_{\text{out}}(\alpha) 
    &= \text{Tr} \left\{ \hat{\rho}_{\text{out}} \left[ \hat{\boldsymbol{a}}\right] \delta(\hat{\boldsymbol{a}} - \boldsymbol{\alpha}) \right\} \\
    &= \text{Tr} \left\{ \hat\rho_{\text{in}}[{\mathcal{T}}^\dagger \hat{\boldsymbol{a}}] \delta(\hat{\boldsymbol{a}} - \boldsymbol{\alpha}) \right\} \\
    &\overset{\hat{\boldsymbol{a}} \rightarrow {\mathcal{T}}  \cdot \hat{\boldsymbol{a}}}{=} \text{Tr} \left\{ \hat{\rho}_{\text{in}} \left[ \boldsymbol{\hat{a}}\right] \delta({\mathcal{T}} \hat{\boldsymbol{a}} - \boldsymbol{\alpha}) \right\} \\
    &\overset{\boldsymbol{\alpha} \rightarrow {\mathcal{T}}^\dagger \boldsymbol{\alpha}}{=} \text{Tr} \left\{ \hat{\rho}_{\text{in}} \left[ \hat{\boldsymbol{a}} \right] \delta(\hat{\boldsymbol{a}} - {\mathcal{T}}^\dagger\boldsymbol{\alpha}) \right\} \\
    &= W_{\text{in}}({\mathcal{T}}^\dagger  \boldsymbol{\alpha}).
\end{aligned}
\end{equation}\par
This input-output relationship from the beam splitter transformation can be regarded as a change of mode basis of the state, defined by $\boldsymbol{\alpha}$. Now consider a joint homodyne measurement $\hat x_i$ and $\hat x_j$ on state $\hat\rho$. The measurement statistics will be written as a joint probability distribution $\mathcal{P}(X_i, X_j)=\langle X_i;X_j| \hat\rho| X_i;X_j\rangle$. With $\alpha_i=X_i+iP_i$, the transformation in Eq.~\ref{eq:wigner} also holds for $X_i$ and $P_i$. Therefore, if the state $\hat\rho$ is transformed through a beam splitter $\hat{\mathcal{U}}_{ij}(\theta)$ applied to its $i$th and $j$th modes, the mixed modes are denoted by $M_{ij}$ and $N_{ij}$. The homodyne measurement statistics of the transformed state is given by
\begin{align}\nonumber
    \mathcal{P}(X_{M_{ij}}, X_{N_{ij}})&=\langle X_i;X_j| \hat{\mathcal{U}}_{ij}\hat\rho\hat{\mathcal{U}}^\dagger_{ij}| X_i;X_j\rangle\\
    &=\langle X_i\text{cos}\theta+X_j\text{sin}\theta;-X_i\text{sin}\theta+X_j\text{cos}\theta| \hat\rho|X_i\text{cos}\theta+X_j\text{sin}\theta;-X_i\text{sin}\theta+X_j\text{cos}\theta\rangle
\end{align}

\subsection{Tripartite joint probability distributions}\label{tri-dist}
When studying quantum properties via deep learning, the choice of the input feature is crucial. Especially in continuous variable systems, where information is encoded in an infinite-dimensional Hilbert space and the space expands exponentially with the number of modes, finding a reasonable input that balances the training complexity with the ability to capture enough correlation information is essential to achieve accurate learning outcomes. 
In our work, instead of feeding the whole density matrix, we select quadrature statistics of multipartite states as the input feature. 
For a tripartite state, the quadrature observables $\hat x_{l}$ and $\hat p_{l}$ in the modes $l= A, B, C$ can be defined as the real and imaginary parts of the photon annihilation operator $\hat a_l$ such that $\hat x_{l} \equiv (\hat a_l + \hat a_l^{\dag})$ and $\hat p_{l} \equiv i(\hat a_l^{\dag} - \hat a_l)$. Homodyne detection then corresponds to a projective measurement of the eigenstates of these quadrature operators. Hence, we define these eigenstates as $\hat x_l| X_l\rangle = X_l | X_l\rangle$ and $\hat p_l| P_l\rangle = P_l | P_l\rangle$, where $X_l$ and $P_l$ describe the continuum of real measurement outcomes for the quadrature measurements in the mode $l$.\par
\begin{figure}[H]
    \centering
    \includegraphics[width=0.55\linewidth]{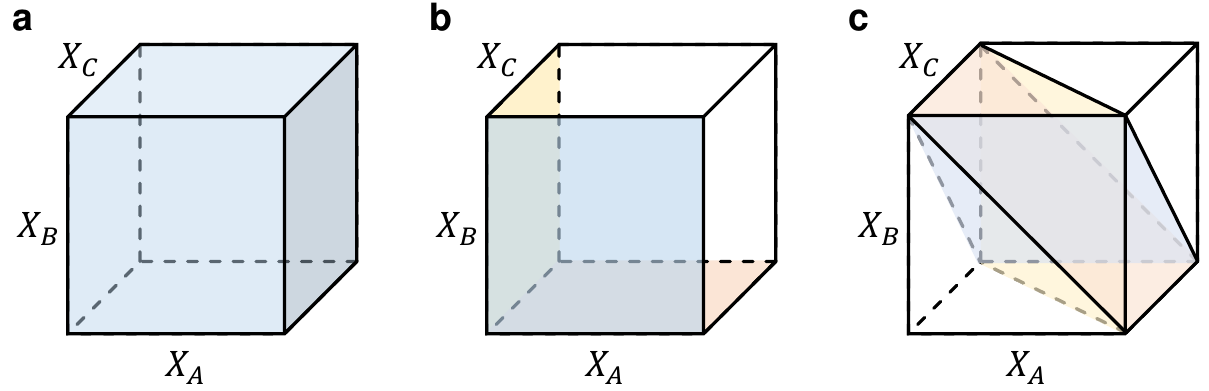}
    \caption{\textbf{Schematic diagram of joint probability distributions of a tripartite quantum state.} \textbf{a,}~Joint probability distribution $\langle X_A;X_B;X_C|\hat\rho| X_A;X_B;X_C\rangle$ in the $(X_A,X_B,X_C)-$space. \textbf{b,}~Three joint probability distributions $\langle X_i;X_j| \hat\rho| X_i;X_j\rangle$ in the $(X_i,X_j)-$plane, where $(i,j)\in\{(A,B),(B,C),(A,C)\}$. \textbf{c,}~Three joint probability distributions $\langle X_i;X_j| \hat\rho| X_i;X_j\rangle$ in the $(X_i,X_j)-$plane, where $(i,j)\in\{(A,M_{BC}),(M_{AC},B),(M_{AB},C)\}$. $M_{ij}$ denotes the first output mode after mixing modes $i$ and $j$ on a balanced beam splitter.}
    \label{fig:prob}
\end{figure}
In Fig.~\ref{fig:prob}, we plot schematic diagrams of three different ways to obtain joint measurement statistics on a tripartite quantum state $\hat\rho$. Figure.~\ref{fig:prob}(a) shows a joint distribution $\mathcal{P}(X_A,X_B,X_C)$ with three variables, representing the measurement statistics of $\langle X_A;X_B;X_C| \hat\rho| X_A;X_B;X_C\rangle$ on three quadratures. This distribution contains the correlation information among quadrature $\hat x_A$, $\hat x_B$, and $\hat x_C$, but the data spans three dimensions, which can be challenging for the network to process as the number of modes increases. 
Therefore, we first consider three marginal distributions of $\mathcal{P}(X_A,X_B,X_C)$, representing joint measurement statistics $\langle X_i;X_j| \hat\rho| X_i;X_j\rangle$ between any pairs $(i,j)\in\{(A,B),(B,C),(A,C)\}$. 
However, these statistics capture only reduced pairwise correlations between modes $i$ and $j$, rather than correlations across bipartitions (see Fig.~\ref{fig:prob}(b)). 
For some states, such as GHZ-type states, which hold genuine multipartite entanglement but do not share entanglement between reduced pairs, these marginal distributions will not accurately capture the useful correlation information.\par
\begin{figure}[H]
    \centering
    \includegraphics[width=0.55\linewidth]{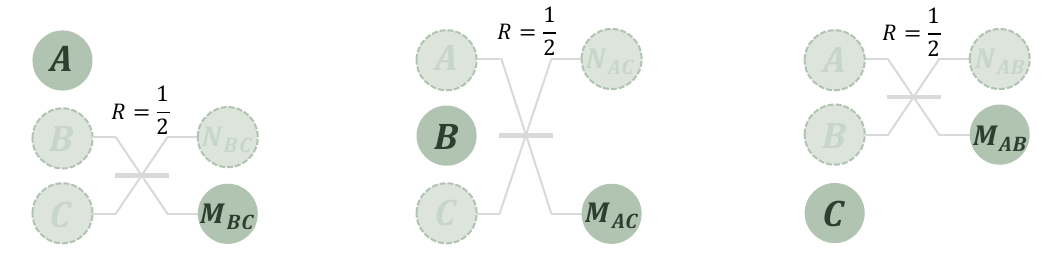}
    \caption{The mixing process of tripartite states.}
    \label{fig:inputs-3mode}
\end{figure}
Consequently, we use the joint measurement statistics after mixing mode pairs $(i,j)\in\{(A,B),(B,C),(A,C)\}$ of the tripartite state on a balanced beam splitter $\hat {\mathcal{U}}_{ij}=\text{e}^{\frac{\pi}{4}(\hat a_i^\dagger\hat a_j-\hat a_i\hat a_j^\dagger)}$.
Mixing modes $i$ and $j$ produces two output modes $M_{ij}$ and $N_{ij}$.
With the input-output relations of the beam splitter, the joint probability distribution after measuring quadratures $\hat x_i$ and $\hat x_{M_{jk}}$ is equal to $\mathcal{P}(X_i,X_{M_{jk}})=\langle X_i;\frac{X_j+X_k}{\sqrt{2}}| \hat\rho| X_i;\frac{X_j+X_k}{\sqrt{2}}\rangle$ on state $\hat\rho$ (see Fig.~\ref{fig:prob}(c)). 

As illustrated in Fig.~\ref{fig:inputs-3mode}, after mixing modes $B$ and $C$, the joint probability distribution in the $(X_A,X_{M_{BC}})-$plane is given by
\begin{align}
    \mathcal{P}(X_A,X_{M_{BC}})=\langle X_A;\frac{X_B+X_C}{\sqrt{2}}| \hat\rho | X_A;\frac{X_B+X_C}{\sqrt{2}}\rangle.
\end{align}\par

After mixing modes $A$ and $C$, the joint probability distribution in the $(X_{M_{AC}},X_B)-$plane is given by
\begin{align}
    \mathcal{P}(X_{M_{AC}},X_B)=\langle \frac{X_A+X_C}{\sqrt{2}};X_B| \hat\rho | \frac{X_A+X_C}{\sqrt{2}};X_B\rangle.
\end{align}\par

After mixing modes $A$ and $B$, the joint probability distribution in the $(X_{M_{AB}},X_C)-$plane is given by
\begin{align}
    \mathcal{P}(X_{M_{AB}},X_C)=\langle \frac{X_A+X_B}{\sqrt{2}};X_C| \hat\rho | \frac{X_A+X_B}{\sqrt{2}};X_C\rangle.
\end{align}

\subsection{Quadripartite joint probability distributions}\label{quadri-dist}
\begin{figure}[hbt]
    \centering
   \includegraphics[width=0.92\linewidth]{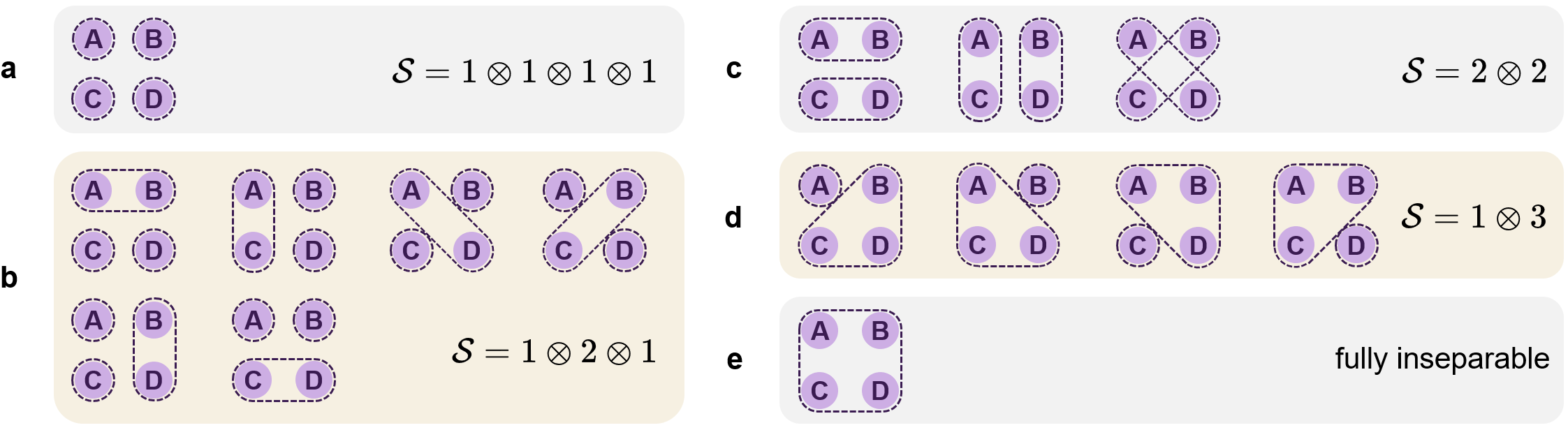}
    \caption{Classification of quadripartite entanglement structures based on the entanglement-structure label $\mathcal{S}$, which represents the multi-partition case of each state. For instance, a quadripartite fully separable state has $\mathcal{S}=1\otimes1\otimes1\otimes1$.}
    \label{fig:ent-4mode}
\end{figure}
\begin{figure}[H]
    \centering
    \includegraphics[width=1\linewidth]{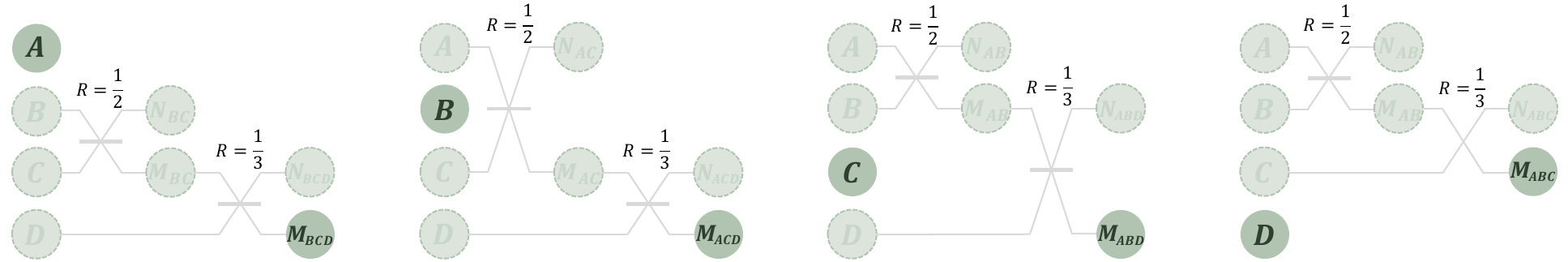}
    \caption{The mixing process of quadripartite states.}
    \label{fig:inputs4mode}
\end{figure}

Figure.~\ref{fig:ent-4mode} demonstrates the quadripartite entanglement structures classified by the entanglement structure label $\mathcal{S}$. The scheme of extracting input features for tripartite states can be easily transferred to states with more quantum modes. In the main text, we select four joint probabilities as features for the quadripartite case. As shown in Fig.~\ref{fig:inputs4mode}, each joint probability is obtained after mixing three modes on two cascaded beam splitters. Here we denote the two output modes by $M_{ijk}$ and $N_{ijk}$ after mixing modes $i$, $j$, and $k$.

After mixing modes $B$, $C$, and $D$, the joint probability distribution in the $(X_A,X_{M_{BCD}})-$plane is given by
\begin{align}
    \mathcal{P}(X_A,X_{M_{BCD}})=\langle X_A;\frac{X_B+X_C+X_D}{\sqrt{3}}| \hat\rho| X_A;\frac{X_B+X_C+X_D}{\sqrt{3}}\rangle.
\end{align}\par

After mixing modes $A$, $C$, and $D$, the joint probability distribution in the $(X_{M_{ACD}},X_B)-$plane is given by
\begin{align}
    \mathcal{P}(X_{M_{ACD}},X_B)=\langle \frac{X_A+X_C+X_D}{\sqrt{3}};X_B| \hat\rho | \frac{X_A+X_C+X_D}{\sqrt{3}};X_B\rangle.
\end{align}\par

After mixing modes $A$, $B$, and $D$, the joint probability distribution in the $(X_{M_{ABD}},X_C)-$plane is given by
\begin{align}
    \mathcal{P}(X_{M_{ABD}},X_C)=\langle \frac{X_A+X_B+X_D}{\sqrt{3}};X_C| \hat\rho | \frac{X_A+X_B+X_D}{\sqrt{3}};X_C\rangle.
\end{align}

After mixing modes $A$, $B$, and $C$, the joint probability distribution in the $(X_{M_{ABC}},X_D)-$plane is given by
\begin{align}
    \mathcal{P}(X_{M_{ABC}},X_D)=\langle \frac{X_A+X_B+X_C}{\sqrt{3}};X_D| \hat\rho | \frac{X_A+X_B+X_C}{\sqrt{3}};X_D\rangle.
\end{align}
\par
\begin{figure}[H]
    \centering   \includegraphics[width=0.80\linewidth]{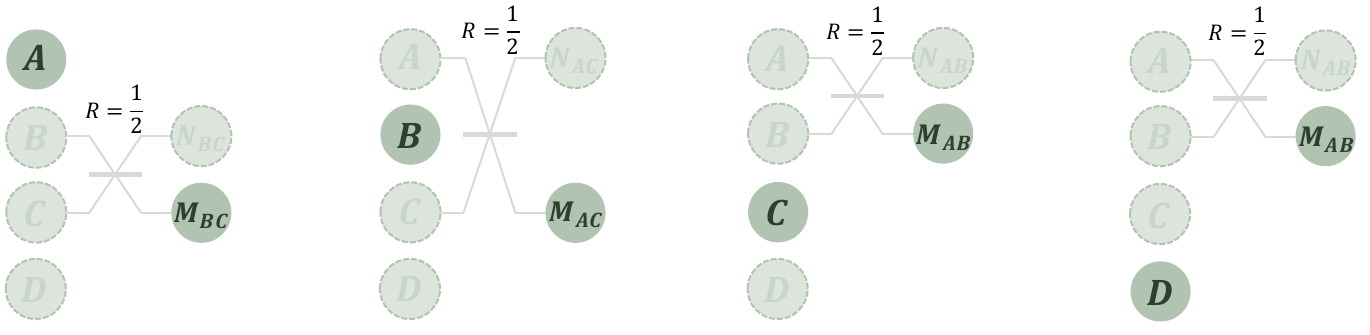}
    \caption{The mixing process of quadripartite states to obtain additional joint probability distributions.}
    \label{fig:inputs4mode-extra}
\end{figure}

The additional quadripartite joint probability distributions are generated after beam splitters mixing according to Fig.~\ref{fig:inputs4mode-extra}.

\subsection{\red{Five-mode joint probability distributions}}
\begin{figure}[hbt]
    \centering
    \includegraphics[width=1\linewidth]{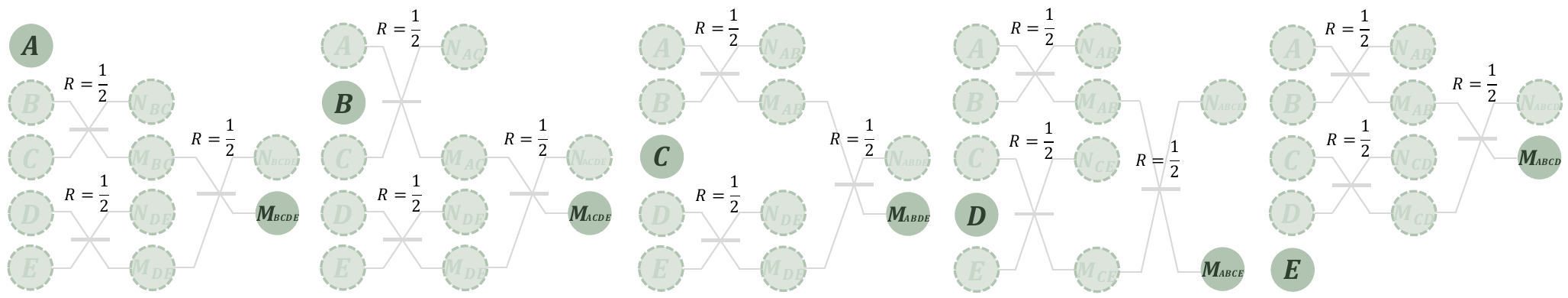}
    \caption{\red{The mixing process of five-mode states.}}
    \label{fig:inputs5mode}
\end{figure}

\red{As shown in Fig.~\ref{fig:inputs5mode}, each joint probability is obtained after mixing four modes on three cascaded beam splitters. Here we denote the two output modes by $M_{ijkl}$ and $N_{ijkl}$ after mixing modes $i$, $j$, $k$, and $l$. For the symmetric mixing used in Fig.~\ref{fig:inputs5mode}, this effective quadrature can be written as}
\begin{equation}
\red{X_{M_{ijkl}}=\frac{X_i+X_j+X_k+X_l}{2}.}
\label{eq:XMijkl}
\end{equation}

\red{Specifically, for the five-mode state $\hat{\rho}$ with modes $A,B,C,D,E$, we use the following five joint probability distributions as input features:}
\begin{align}
\red{\mathcal{P}(X_A, X_{M_{BCDE}})} 
&
\red{= 
\Big\langle X_A;\, \frac{X_B+X_C+X_D+X_E}{2}\Big|\hat{\rho}\Big|X_A;\, \frac{X_B+X_C+X_D+X_E}{2}\Big\rangle,}
\label{eq:S5_1}\\
\red{\mathcal{P}(X_{M_{ACDE}}, X_B)} 
&
\red{=\Big\langle \frac{X_A+X_C+X_D+X_E}{2};\, X_B\Big|\hat{\rho}\Big|\frac{X_A+X_C+X_D+X_E}{2};\, X_B\Big\rangle,}
\label{eq:S5_2}\\
\red{\mathcal{P}(X_{M_{ABDE}}, X_C) }
&
\red{=
\Big\langle \frac{X_A+X_B+X_D+X_E}{2};\, X_C\Big|\hat{\rho}\Big|\frac{X_A+X_B+X_D+X_E}{2};\, X_C\Big\rangle,}
\label{eq:S5_3}\\
\red{\mathcal{P}(X_{M_{ABCE}}, X_D) }
&
\red{=
\Big\langle \frac{X_A+X_B+X_C+X_E}{2};\, X_D\Big|\hat{\rho}\Big|\frac{X_A+X_B+X_C+X_E}{2};\, X_D\Big\rangle,}
\label{eq:S5_4}\\
\red{\mathcal{P}(X_{M_{ABCD}}, X_E)} 
&
\red{=\Big\langle \frac{X_A+X_B+X_C+X_D}{2};\, X_E\Big|\hat{\rho}\Big|\frac{X_A+X_B+X_C+X_D}{2};\, X_E\Big\rangle.}
\label{eq:S5_5}
\end{align}

\subsection{Discretization of correlation patterns}
The correlation pattern of the joint probability distribution $\mathcal{P}(X_i,X_j)$ is defined continuously over the $(X_i,X_j)$ plane, as well as other distributions $\mathcal{P}(X_i,P_j)$, $\mathcal{P}(P_i,X_j)$ and $\mathcal{P}(P_i,P_j)$. 
Directly feeding them into neural networks is impossible, as the model requires limited and discretized data for effective processing. 
Therefore, we restrict the region of phase space from -6 to 6 and bin each distribution image into a normalized $24\times24$-dimensional matrix $M$. 
The matrix element $M_{mn}$ of distribution $\mathcal{P}(X_i,X_j)$ (other joint statistics are processed analogously) is given by its median value in the region $X_i\in(-6+\frac{1}{2}(n-1),-6+\frac{1}{2}n), X_j\in(6-\frac{1}{2}(m-1), 6-\frac{1}{2}m)$ to discretize the joint probability distribution. \par


\section{Neural network details}
\subsection{Neural network architecture}
\begin{figure}[hbt]
    \centering
    \includegraphics[width=0.65\linewidth]{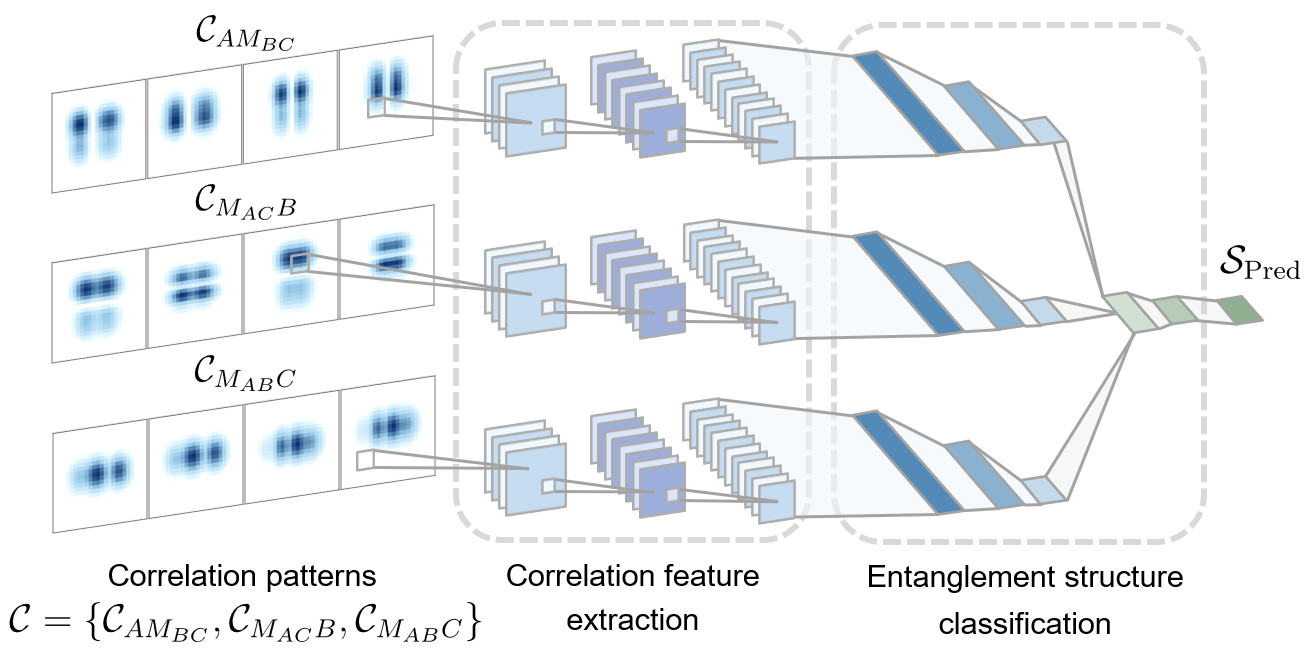}
    \caption{The neural network architecture for tripartite entanglement structures classification.}
    \label{fig:NN}
\end{figure}

The neural network architecture illustrated in Fig.~\ref{fig:NN} is designed to process three distinct sets of input correlation patterns from tripartite states, labeled as $\mathcal{C}_{AM_{BC}},\,\mathcal{C}_{M_{AC}B}$ and $\mathcal{C}_{M_{AB}C}$, each of which has shape of $(24, 24, 4)$. 
For each set, a shared sub-network is employed to extract meaningful features. 
This sub-network is composed of three convolutional layers, where the first layer uses 32 filters, the second layer 64 filters, and the third layer 128 filters, all with $3\times3$ kernels, ReLU activation, and same padding.
These layers are followed by max-pooling operations to downsample the feature maps, effectively capturing the correlation features while reducing the spatial dimensions.
After convolutional layers, the feature maps are flattened into 1D vectors, which are then passed through a series of fully connected layers. 
The fully connected layers consist of 1024, 256, and 8 units, respectively, representing the extracted features from the input correlation patterns.
\par

The outputs from the three sub-networks, corresponding to the three extracted correlation features, are then concatenated into a single feature vector. 
This concatenated vector is processed by a final classification network, which begins with a fully connected layer of 8 units with ReLU activation, followed by another fully connected layer with 3 units, also with ReLU activation. 
The final output layer consists of 3 units with softmax activation, producing a probability distribution over the three possible classes with entanglement structure labels for fully separable partition $\mathcal{S}=1\otimes1\otimes1$, biseparable partition $\mathcal{S}=1\otimes2$, and fully inseparable states, encoded into a one-hot array. 
This architecture is optimized for extracting and integrating correlation features from the experimentally feasible patterns, ultimately enabling the precise classification of the entanglement structures. 
\par
The neural network that processes correlation patterns from quadripartite states is designed the same way as in Fig.~\ref{fig:NN}. The number of sub-networks is increased from three to four for the four correlation pattern sets, and the output layer is a probability distribution vector over the five classes representing states with entanglement structure labels for fully separable partition $\mathcal{S}=1\otimes1\otimes1\otimes1$, tri-separable partition $\mathcal{S}=1\otimes1\otimes2$, biseparable partitions $\mathcal{S}=2\otimes2$ and $\mathcal{S}=1\otimes3$, and fully inseparable states.

\begin{figure}[bht]
    \centering
    \includegraphics[width=0.95\linewidth]{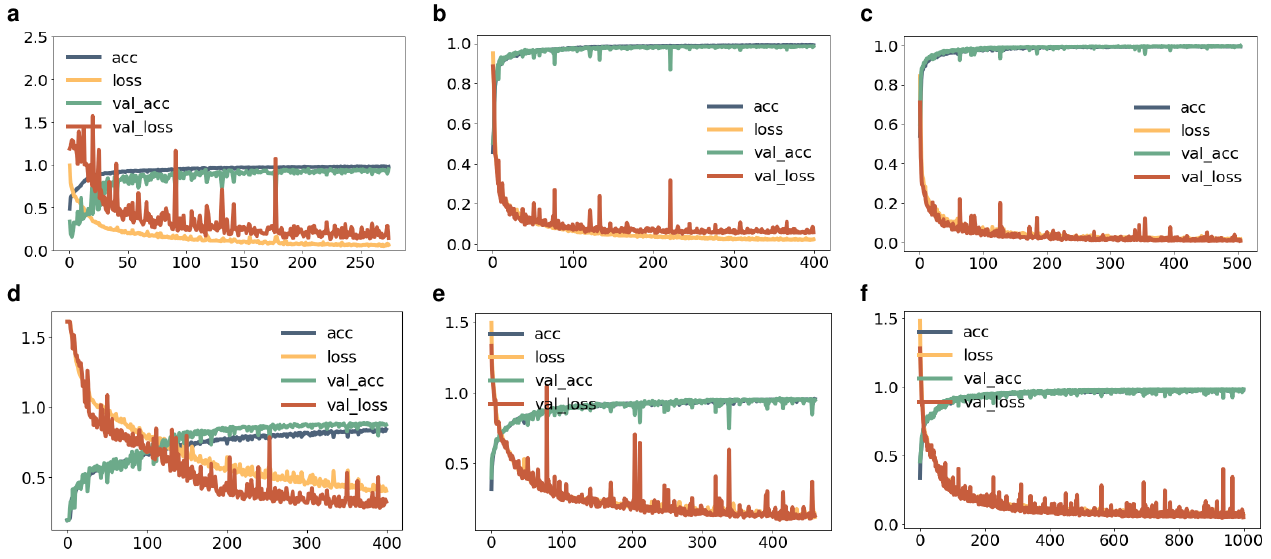}
    \caption{Accuracies and losses (categorical cross entropy) during the training process on the training dataset (blue lines for accuracy and yellow lines for loss) and validation dataset (green lines for accuracy and red lines for loss), which consist of both finite- and infinite-stellar rank states. The training process is stopped when the validation loss converges. (a)~Training plots for the original tripartite dataset with $24\,000$ samples. There is a notable divergence that emerges between the training and validation losses, where the training loss tends to zero, indicating the model fits the training data well. In contrast, the validation loss has increased slightly, indicating that the model's performance on unseen data is declining. This is called the overfitting problem.
    (b)~Training plots for four-times-augmented tripartite dataset with $96\,000$ samples. 
    (c)~Training plots for six-times-augmented tripartite dataset with $144\,000$ samples. The model generalizes better to unseen data, effectively mitigating the overfitting observed in (a). 
    (d)~Training plots for the original quadripartite dataset with $20\,000$ samples. 
    (e)~Training plots for twelve-times-augmented quadripartite dataset with $240\,000$ samples. 
    (f)~Training plots for 24-times-augmented quadripartite dataset with $480\,000$ samples. The model generalizes better to unseen data, effectively mitigating the overfitting observed in (d).}
    \label{fig:trainprocess}
\end{figure}

\subsection{Neural network training process}
In Fig.~\ref{fig:trainprocess}, we compare the training process of neural networks trained on the original and augmented datasets. 
During the training process, $70\%$ of the training datasets are used for updating the model's parameters, and the rest are for validation used to evaluate the model's performance and tune hyperparameters.
The plots illustrate the training and validation metrics for various dataset configurations.
\par

Figures (a), (b), and (c) illustrate the model's performance on the original tripartite dataset with $24\,000$ samples, and its four-times and six-times augmented versions with $96\,000$ and $144\,000$ samples, respectively. 
Figures (d), (e), and (f) correspond to the original quadripartite dataset with $20\,000$ samples, and its twelve-times and 24-times augmented versions with $240\,000$ and $480\,000$ samples, respectively.
The results indicate that with smaller original datasets, such as in (a) and (d), the model tends to overfit, as demonstrated by a growing divergence between training and validation losses. 
The training loss continues to decrease toward zero, reflecting that the model is effectively memorizing the training data. 
However, this comes at the cost of generalization, as evidenced by the slight increase in validation loss, indicating the model's diminishing ability to perform well on unseen data.
\par

As the dataset size increases through the QDA method, especially in (c) and (f), the validation performance improves, the gap between training and validation losses decreases, and the model generalizes better to unseen data.
Therefore, the QDA method effectively mitigates the overfitting problem.

\subsection{Ablation study for different QDA operations}
In the tripartite case, given the scarcity of fully separable states (1,600 samples out of 17,600 total training samples), the neural network demonstrates limited predictive performance, with an accuracy of 93.2\%. Then we expand the set of fully separable states to 8,000 samples through three different strategies, yielding 24,000 total training samples. By applying only mode permutation, the accuracy increases to 96.5\%. When employing only convex combinations, the accuracy reaches 96.0\%. With both operations combined, the model achieves 96.4\%. These results indicate that different QDA operations lead to slightly different network performance.

\subsection{Calculation time of direct simulation and data augmentation}
As we show in the Results in the main text, generating the initial dataset through simulation is considerably more time-consuming than producing new samples via data augmentation, as evidenced by the comparison in Table.~\ref{tab:generation_time}. The time required to simulate a single new sample is significantly greater than the time needed to generate one through augmentation ($\sim10^6$). 
Moreover, simulating quantum states becomes increasingly challenging as the number of modes grows, further amplifying the computational complexity and resource demands. 
In contrast, data augmentation provides an efficient and scalable alternative, allowing for the rapid expansion of the dataset.

\begin{table}[H]
    \centering
    \caption{Average time required to generate a sample (with finite or infinite stellar rank) through simulation and data augmentation. The simulation process involves generating density matrices, producing correlation patterns, and calculating the entanglement labels via the Quantum Fisher Information (QFI) criterion.}
    \label{tab:generation_time}
    \renewcommand{\arraystretch}{1.0}
    \setlength{\tabcolsep}{4pt}
    \begin{tabular}{lcc}
        \hline\hline
        \multirow{2}{*}{\textbf{Method}}& \multicolumn{2}{c}{\textbf{Per sample (s)}} \\
        \cmidrule{2-3}
         & \textbf{Tripartite} & \textbf{Quadripartite} \\
        \hline
        Direct simulation                 & 83.24          & 342.47 \\
        QDA: Mode Permutation    & $6.33 \times 10^{-5}$ & $2.51 \times 10^{-4}$ \\
        QDA: Convex Combination  & $2.33 \times 10^{-4}$ & $7.52 \times 10^{-4}$ \\
        \hline\hline
    \end{tabular}
\end{table}

\subsection{Fine-grained classification of tripartite entanglement structures}
We perform an additional test on $8\,000$ tripartite samples. 
Using the same network architecture and input correlation patterns, we reconfigured the output layer to a one-hot encoding of five specific entanglement configurations: fully separable ($A|B|C$), three types of biseparable ($A|BC$, $AC|B$, $AB|C$), and fully inseparable ($ABC$) states. 
With sixfold QDA, the model's accuracy improved from 70.8\% (without QDA) to 94.1\%.

\bibliographystyle{sn-nature}

\end{document}